\documentclass[journal]{vgtc}                     

\onlineid{4241}

\vgtccategory{Research}

\vgtcpapertype{Application}

\title{\systemname: Visual Analytics of Human Mobility Data for Understanding and Mitigating Urban Segregation}

\author{%
  \authororcid{Yue Yu}{0000-0002-1825-0097},
  \authororcid{Yifang Wang}{0000-0001-6267-9440}, \authororcid{Yongjun Zhang}{0000-0002-8265-925X}, \authororcid{Huamin Qu}{000-0002-3344-9694}, and 
  \authororcid{Dongyu Liu}{0000-0002-8915-2785}
}

\authorfooter{
  \item
  	Yue Yu and Huamin Qu are with The Hong Kong University of Science and Technology.
  	E-mail: yue.yu@connect.ust.hk, huamin@cse.ust.hk.
  \item
  	Yifang Wang is with The Center for Science of Science and Innovation, Northwestern University. E-mail: yifang.wang@kellogg.northwestern.edu.
  \item Yongjun Zhang is with Stony Brook University. Email: yongjun.zhang@stonybrook.edu.
  \item Dongyu Liu is with the University of California, Davis and is the corresponding author. E-mail: dyuliu@ucdavis.edu.
}

\abstract{
Urban segregation refers to the physical and social division of people, often driving inequalities within cities and exacerbating socioeconomic and racial tensions.
While most studies focus on residential spaces, they often neglect segregation across ``activity spaces'' where people work, socialize, and engage in leisure. 
Human mobility data offers new opportunities to analyze broader segregation patterns, encompassing both residential and activity spaces, but challenges existing methods in capturing the complexity and local nuances of urban segregation.
This work introduces \systemname, a novel visual analytics system for multi-level analysis of urban segregation, facilitating the development of targeted, data-driven interventions.
Specifically, we developed a deep learning model to predict mobility patterns across social groups using environmental features, augmented with explainable AI to reveal how these features influence segregation.
The system integrates innovative visualizations that allow users to explore segregation patterns from broad overviews to fine-grained detail and evaluate urban planning interventions with real-time feedback.
We conducted a quantitative evaluation to validate the model's accuracy and efficiency. Two case studies and expert interviews with social scientists and urban analysts demonstrated the system's effectiveness, highlighting its potential to guide urban planning toward more inclusive cities.
}


\graphicspath{{figs/}{figures/}{pictures/}{images/}{./}} 
\usepackage{hanging}
\usepackage{enumitem}
\usepackage{tabu}                      
\usepackage{booktabs}                  
\usepackage{lipsum}                    
\usepackage{mwe}                       
\usepackage{mathptmx}                  
\usepackage{color}
\usepackage[dvipsnames,svgnames]{xcolor}
\usepackage{tcolorbox}
\usepackage{listings}
\usepackage{multirow}
\usepackage{float}

\definecolor{wyftextcolor}{RGB}{237, 85, 106} 

\definecolor{yytextcolor}{RGB}{0, 0, 0} 

\definecolor{dypink}{HTML}{ec008c}

\newcommand{\yy}[1]{\textcolor{yytextcolor}{#1}}

\usepackage{balance}
\usepackage{soul}

\newcommand\eg{e.g.}
\newcommand\et{et al.}
\newcommand{\systemname}{\textit{InclusiViz}\xspace}

\definecolor{commColor}{HTML}{6082B3}
\definecolor{commLightColor}{HTML}{E2EBF9}
\definecolor{cbgColor}{HTML}{B59872}
\definecolor{cbgLightColor}{HTML}{FFEFDA}
\definecolor{poiColor}{HTML}{AC757E}
\definecolor{poiLightColor}{HTML}{F1D5D9}

\newcommand\commpill{\textcolor{commColor}{\textbf{Comm}}}
\newcommand\cbgpill{\textcolor{cbgColor}{\textbf{CBG}}}
\newcommand\poipill{\textcolor{poiColor}{\textbf{POI}}}

\newcommand\TIpill{\tcbox[on line, colframe=commColor, colback=commLightColor,coltext=black,
arc=0.5pt, boxsep=0mm, left=1.8pt, right=1.8pt, top=1pt, bottom=1pt, boxrule=0.3mm]{\textbf{R1}}}
\newcommand\TIIpill{\tcbox[on line, colframe=commColor, colback=commLightColor,coltext=black,
arc=0.5pt, boxsep=0mm, left=1.8pt, right=1.8pt, top=1pt, bottom=1pt, boxrule=0.3mm]{\textbf{R2}}}
\newcommand\TIIIpill{\tcbox[on line, colframe=cbgColor, colback=cbgLightColor,coltext=black,
arc=0.5pt, boxsep=0mm, left=1.8pt, right=1.8pt, top=1pt, bottom=1pt, boxrule=0.3mm]{\textbf{R3}}}
\newcommand\TIVpill{\tcbox[on line, colframe=cbgColor, colback=cbgLightColor,coltext=black,
arc=0.5pt, boxsep=0mm, left=1.8pt, right=1.8pt, top=1pt, bottom=1pt, boxrule=0.3mm]{\textbf{R4}}}
\newcommand\TVpill{\tcbox[on line, colframe=poiColor, colback=poiLightColor,coltext=black,
arc=0.5pt, boxsep=0mm, left=1.8pt, right=1.8pt, top=1pt, bottom=1pt, boxrule=0.3mm]{\textbf{R5}}}

\newcommand{\commview}{\textit{Community View}\xspace}
\newcommand{\cbgview}{\textit{CBG View}\xspace}
\newcommand{\mapview}{\textit{Map View}\xspace}
\newcommand{\whatifview}{\textit{What-If View}\xspace}

\newcommand{\flowmat}{\textit{Flow Matrix}\xspace}
\newcommand{\commsig}{\textit{Community Signature}\xspace}
\newcommand{\cbgTable}{\textit{Ranking Table}\xspace}

\newcommand{\featControl}{\textit{Feature Control Panel}\xspace}
\newcommand{\featImpact}{\textit{Feature Impact Plot}\xspace}
\newcommand{\featImpacts}{\textit{Feature Impact Plots}\xspace}
\newcommand{\bubbleglyph}{\textit{Filter Bubble Glyph}\xspace}
\newcommand{\realmap}{\textit{Navigational Map}\xspace}
\newcommand{\dorlingmap}{\textit{Dorling Map}\xspace}

\newcommand{\parasum}[1]{\vspace*{-0.1cm}\paragraph{\textbf{\footnotesize #1}}}

\begin{document}



\maketitle

\section{Introduction}

Imagine invisible lines, not physical walls, dividing a city; residents are segregated based on their wealth, race, and political beliefs. People live in bubbles and rarely interact with those from different sociodemographic backgrounds.
This form of division, known as \textbf{urban segregation}, transcends geography and reinforces social inequalities.
To reshape cities into more inclusive and equitable spaces, social scientists and policymakers have increasingly turned their focus to understanding these complex and invisible barriers and attempting to propose actionable interventions to dismantle these divisions~\cite {chetty2020income, nilforoshan2023human}.

Traditionally, research on urban segregation has concentrated on residential areas, utilizing census statistics such as racial composition
to analyze segregation patterns~\cite{white1986segregation}.
While informative, this approach provides a narrow lens that only focuses on residential environments. 
It overlooks the full picture of daily activities beyond residential areas, such as workplaces and social venues, potentially biasing the analysis. 
The recent availability of GPS-based human mobility data is broadening segregation studies from ``residential space'' to ``activity space''~\cite{cagney2020urban},  offering a comprehensive view of how people experience segregation in their daily routine~\cite{wong2011measuring, browning2017socioeconomic, zhang2023human}. 
For instance, researchers try to explain the motivations behind mobility segregation by indexing the segregation and modeling the human daily activity trajectories~\cite{moro2021mobility} and suggest urban planning interventions to bridge 
segregated communities~\cite{nilforoshan2023human}.

However, existing GPS-based approaches primarily rely on data aggregation to offer a general overview of urban segregation, falling short in capturing detailed patterns of local interactions between neighborhoods at a finer scale.
Moreover, these methods lack \textit{data-informed} tools to guide decision-making and \textit{feedback-driven} processes to adjust strategies, making it difficult for social scientists and policymakers to evaluate urban planning interventions effectively. 
Visual analytics (VA) is an ideal approach to overcome these limitations, but designing an effective VA system in this context is challenging.

First, a city's mobility dataset typically comprises thousands of spatial units with potential movements between any two.
Aggregation techniques help grasp mobility networks~\cite{von2015mobilitygraphs, zhou2018visual, liu2019tpflow}, but deriving actionable, fine-grained insights for local planning remains a challenge.
Several VA systems have shown promise in urban planning contexts, such as optimizing location choices~\cite{liu2016smartadp, chen2023fslens} or enhancing bus network efficiency~\cite{weng2020towards}. However, they are not tailored for segregation analysis, which requires navigating the socio-spatial complexities of segregation. 
Second, understanding intricate correlations between environmental features and mobility patterns is essential for designing effective interventions against segregation. 
These correlations are often nonlinear, multifaceted, and vary across regions~\cite{athey2021estimating}.
Existing systems can effectively visualize the mobility flows between areas~\cite{adrienko2010spatial, andrienko2016revealing} or examine environmental impacts within specific regions~\cite{lei2023geoexplainer, yu2023neighviz}, but they cannot predict or explain mobility flows in relation to segregation patterns.

To address these challenges, we propose \systemname, a visual analytics system for understanding urban segregation using mobility data, effectively meeting complex segregation analysis demands.
First, to support fine-grained analysis of segregation patterns, we introduce a three-level workflow guiding users from citywide overview to neighborhood-specific insights.
We designed metaphor-based bubble glyphs in a Dorling layout to help users interpret neighborhood interactions efficiently.
Second, to unravel intricate correlations between environmental features and mobility patterns, we developed a deep learning model based on Deep Gravity~\cite{simini2021deep} to predict mobility flows using these features. The model is equipped with feature impact visualizations, offering instance-level insights by illustrating how each feature contributes to visitor flows from different social groups.
Finally, to bridge insights and action, \systemname features a what-if functionality leveraging the model, allowing users to test urban planning interventions and receive immediate feedback on their effects on segregation patterns.


In summary, the main contributions of this work are as follows:
\begin{itemize}[topsep=0pt, itemsep=0.1pt]
    \item
    We systematically delineate the problem domain for understanding and mitigating urban segregation using mobility data and present three levels of socio-spatial analysis.
    
    \item 
    We introduce a comprehensive data analysis pipeline that supports mobility-based community detection, segregation analysis, mobility prediction feature impact explanations, and what-if analyses to guide precise urban planning interventions.
    
    \item 
    We develop \systemname, a visual analytics system with novel designs, such as metaphor-based bubble glyphs for interactively exploring segregation and evaluating planning scenarios.

    \item 
    We evaluate \systemname with two case studies and interviews with domain experts to show the system's effectiveness and usability. 
\end{itemize}

\section{Related Works}

\subsection{Segregation Analysis in Mobility Data}
\label{sec:related_segregation}
Two approaches are typically applied by social scientists to analyze segregation in mobility data: (1) the index-based approach, which measures the segregation level, and (2) the model-based approach, which uncovers underlying reasons for the segregation.

\textbf{Segregation index} 
is a straightforward approach as an indicator of understanding the segregation level in human mobility data.
The index is usually computed based on the demographics of the group of people who visit a place during a period of time. 
The computation formula of the segregation index is diverse, which can be information entropy~\cite{huang2022unfolding}, isolation index~\cite{white1986segregation, athey2021estimating}, or just distribution unevenness~\cite{zhang2023human, moro2021mobility}. 
Depending on the research topic, the main subject of the segregation can involve different attributes, including income~\cite{cheng2025economic}, partisan~\cite{zhang2023human}, race~\cite{athey2021estimating}, or a combination of them~\cite{huang2022unfolding}.
\yy{
To better communicate these patterns, various visualization techniques have been developed, from traditional segregation curves~\cite{duncan1955segregation} for two-group cases to more recent methods like segplots~\cite{elbers2024segplot} that can display multi-group distributions within spatial units.}
However, the segregation index \yy{and its visualizations}, being merely an aggregated descriptive metric, does not explain the underlying reasons for the segregation patterns in mobility data.

\textbf{Mobility models} offer researchers a detailed examination of the features influencing mobility and segregation patterns.
A classic model is Exploration and Preferential Return (EPR)~\cite{song2010modelling, pappalardo2015returners}, which uses two key parameters, exploration (visiting new places) and preferential return (revisiting previously visited locations), to model the dynamics of human movement trajectories.
Moro \et~\cite{moro2021mobility} extend the EPR by adding a parameter that quantifies an individual motivation to visit new places where their income group is in the minority. 
Nevertheless, modeling complex human behaviors with few parameters may oversimplify the complexity of mobility patterns and their driving factors.
More sophisticated models, like Deep Gravity~\cite{simini2021deep}, utilize a comprehensive set of features—including facility density and demographic information—and employ deep neural networks to predict mobility flow probabilities between locations.
Despite their predictive power, the complexity of neural networks trained on extensive datasets can obscure the key features influencing segregation patterns, making it challenging to derive actionable insights for mitigating segregation.

We propose a data analysis pipeline that computes the segregation indices and models the impact of environmental features on mobility using an adapted Deep Gravity model, with additional support for feature impact explanations and what-if analyses.

\subsection{Human Mobility Visualization}
Human mobility analysis has been widely studied in the visualization community~\cite{deng2023survey}. 
Here, we mainly discuss the most relevant literature about mobility data visualization and mobility model visualization.

\textbf{Mobility data visualization} often uses origin-destination (OD) flow map~\cite{guo2014origin,NI201757} to visualize an individual’s movement trajectories.
However, OD flow maps can become visually cluttered with dense data. To address this, researchers have developed alternative visualization techniques such as matrix representations, where each cell indicates flow volume while preserving spatial layout~\cite{wood2010visualisation, yang2016many}. 
Additionally, techniques leveraging data mining, including clustering~\cite{von2015mobilitygraphs, liu2019tpflow, shi2020urbanmotion} and Word2Vec~\cite{zhou2018visual}, have been employed to aggregate mobility data and distill key insights, thereby enhancing visualization clarity and interpretability.
Other mobility visualization work further incorporates semantic information, such as point of interest (POI) information~\cite{zeng2017visualizing,LI201850} and social media text~\cite{chen2015interactive, krueger2019bird}, enriching the contextual understanding of mobility patterns.
However, a notable gap in the literature is the insufficient integration of demographic data into mobility visualization, which is essential for understanding underlying social dynamics such as income and racial inequality~\cite{athey2021estimating, moro2021mobility} and partisan polarization~\cite{zhang2023human}.

\textbf{Mobility model visualization}, primarily studied in the transportation domain, helps understand the driving factors of mobility patterns.
For instance, visualizing congestion predictions generated by deep learning models can assist experts in analyzing and predicting congestion scenarios~\cite{lee2019visual} and even refining the models~\cite{jin2022visual}.
Meanwhile, TCEVis~\cite{dong2023tcevis} employs SHAP values to explain features influencing traffic congestion.
However, these models typically focus on entire populations without considering variations across social groups.

Our work not only visualizes the mobility patterns of diverse social groups but also incorporates these distinctions into mobility modeling. 
This approach allows for a visual comparison of the factors driving segregation among different social groups, enhancing the understanding of mobility dynamics within a sociodemographic context.

\subsection{What-If Analysis}
What-if analysis empowers users to explore the predictive models by adjusting input features, thereby understanding the model's behavior under various hypothetical scenarios.
Visual analytics systems enhance this exploration by facilitating interactive feature adjustments to observe their impact on predictions~\cite{krause2016interacting, wexler2019if, cheng2021vbridge}.
In the urban planning scenario, what-if analysis is widely embedded in many planning support systems, such as $\rm{What-If?}^{TM}$~\cite{klosterman2008whatif} and CommunityViz~\cite{janes2008communityviz}, as well as traffic analysis systems~\cite{andrienko2016leveraging} and public policy simulators~\cite{yang2022epimob, usman2022multiscale}. 
These systems facilitate data-driven decision-making by enabling stakeholders to simulate and evaluate the consequences of various urban design scenarios. 
However, a common limitation among these tools is their lack of focus on the goal of promoting inclusivity through urban planning.
CityScope~\cite{alonso2018cityscope} more closely aligns with our objectives by offering a tangible interface that promotes urban performance metrics, including the exposure diversity of different social groups.
Nevertheless, the agent-based modeling behind CityScope, relying on parameters distilled from surveys or regression~\cite{grignard2018impact}, may oversimplify the diverse social and environmental nuances of urban settings.

Our approach addresses these limitations by developing a deep learning model that not only captures complex correlations between environmental features and segregation but enables a what-if analysis to provide real-time feedback on the effectiveness of user-defined interventions for segregation mitigation. We further presented a set of tailored visualizations based on superposition approaches to support comparative analysis between different intervention strategies.
\section{Background}


\subsection{Requirement Analysis}
\label{prob_formulation}
\label{sec:03_Background_VisualizationDesignTasks} 

Over the past year, during the design phase of our system, we closely collaborated with social scientist $E_A$ to shape the design of \systemname.
$E_A$ is an experienced sociologist in using human mobility data for urban segregation analysis with a number of relevant publications focusing on the causes and mitigation of partisan segregation. 
During the evaluation phase, we also involved experts $E_B$, $E_C$, and $E_D$, all with a focus on segregation but from diverse backgrounds (detailed in Section~\ref{sec:evaluation}).
Through monthly discussions with $E_A$ and extensive literature reviews, we found that traditional urban segregation studies mainly focus on designing segregation indices and using regression analyses to associate environmental features with segregation levels~\cite{zhang2023human, huang2022unfolding, athey2021estimating}.

These methods offer an overview of segregation patterns, but examining detailed local mobility dynamics and the interactions between spatial units often requires labor-intensive, ad-hoc analysis, which can be time-consuming.
To address these limitations, we have first identified two primary analysis goals, \textit{Community-level Analysis} and \textit{Census Block Group (CBG)-level Analysis}, to facilitate a comprehensive exploration and analysis of segregation patterns from community-level overview to detailed local dynamics among different social groups.
In our study, a CBG is a basic and precisely defined spatial unit with 600-3000 residents; a community consists of multiple CBGs, resembling a larger district within a city.
``Social group'' refers to distinct categories of individuals sharing common characteristics, such as income, race, partisan affiliation, or other demographic attributes.

During our discussions, $E_A$ expressed keen interest in identifying strategic urban planning methods to counteract the observed segregation, alongside a straightforward approach to assess the impact of proposed plans.
Though there are tools for simulating the effects of urban policies~\cite{klosterman2008whatif, janes2008communityviz, alonso2018cityscope}, he noted they lack integration with segregation analysis, making it challenging to assess the social impacts of urban planning.
Also, among various urban planning interventions, $E_A$ suggested adjusting different types of Points of Interest (POIs) as a practical strategy for addressing segregation.
Consequently, we identified the third goal: \textit{Point of Interest (POI)-level Analysis}, to allow for interactive exploration and assessment of POI modifications to promote inclusivity.
Overall, we have summarized the following three-level socio-spatial analysis goals.

\textbf{\commpill unity-level Analysis} enables experts to obtain an overview of the city’s segregation dynamics at the community level. This granularity helps identify focus areas for detailed analysis by examining mobility patterns and sociodemographic characteristics of communities.


\textbf{\cbgpill-level Analysis} facilitates the identification of specific CBGs that could serve as pilot areas for mitigating segregation within selected communities. This involves efficiently pinpointing CBGs with significant segregation but intervention potential and exploring how their features influence the movement of people from different social groups.


\textbf{\poipill-level Analysis} allows experts to adjust POIs within selected CBGs to reduce segregation and promote inclusivity. This involves comparing the effectiveness of different POI modifications in fostering inclusive environments.

We have further identified five design requirements to support the three-level socio-spatial analysis goals: 

\hangpara{2em}{1}
\TIpill{} \textbf{Segment and characterize communities.} 
The system should identify communities based on human mobility and characterize them by the demographics of population flows, enabling the expert to understand how populations move between communities and how citywide segregation is shaped by these flows.

\hangpara{2em}{1}
\TIIpill{} \textbf{Compare sociodemographics across communities.} 
The system should support a comparative analysis of sociodemographic attributes across communities, including both residents and visitors. 
This will enable the expert to detect communities where segregation in activity spaces is prominent, guiding further investigation.

\hangpara{2em}{1}
\TIIIpill{} \textbf{Identify CBGs with intervention potential.} 
Within a selected community, the system should assist the expert in identifying CBGs that show significant segregation levels and high potential for effective interventions. 
A multi-criteria ranking of segregation-related metrics of all CBGs within this community is thus necessary to prioritize areas for targeted action. 

\hangpara{2em}{1}
\TIVpill{} \textbf{Analyze target CBG inflow and influential features.} 
Once a target CBG is selected, the system should provide a detailed analysis of the inflow patterns across social groups and illustrate how key environmental features influence these patterns.
This will help the expert identify factors contributing to segregation. 
By distinguishing how each social group responds to these features, the expert can better assess if certain groups are excluded or isolated, offering deeper insights into segregation dynamics.

\hangpara{2em}{1}
\TVpill{} \textbf{Simulate urban planning interventions.} 
To facilitate practical applications, the system should support the expert in conducting ``what-if'' analyses by modifying POI features in the target CBG to reduce segregation.
This requires simulating and visualizing the real-time impact of these changes on mobility patterns and segregation levels, supporting the expert to assess the practical interventions for promoting inclusivity.

\begin{figure} []
 \centering 
 \vspace{-0.2cm}
 \includegraphics[width=\linewidth]{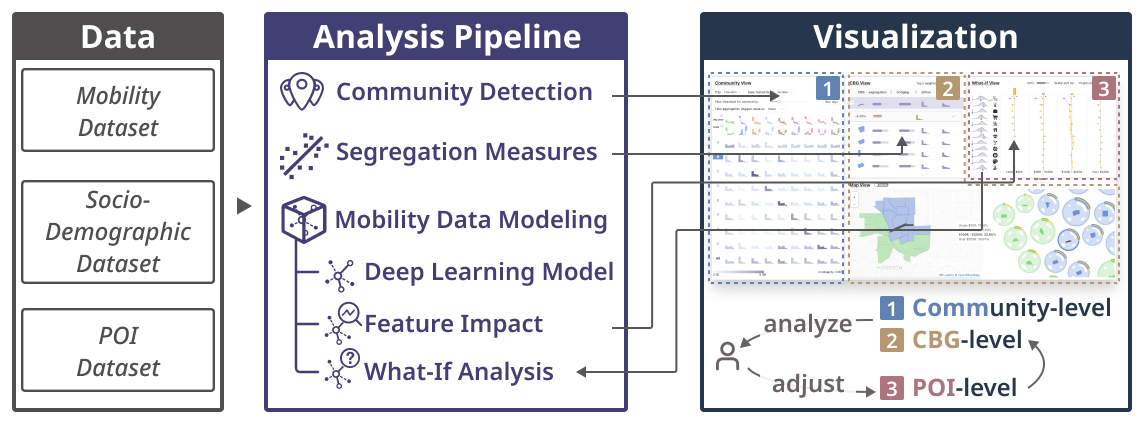}
 \vspace{-0.5cm}
 \caption{
    System overview - \systemname comprises three modules: Data, Analysis Pipeline, and Visualization, to support interactive segregation analysis (\commpill{} and \cbgpill) and iterative intervention development (\poipill).
 } 
 \vspace{-0.7cm}
 \label{fig:system-overview}
\end{figure}

\subsection{System Overview}
\label{sec:system-overview} 
To address the identified requirements, we developed \systemname, an open-source human-in-the-loop VA system\footnote{\url{https://github.com/bruceyyu/inclusiviz/}} with three modules (Fig.~\ref{fig:system-overview}).
The Data module stores data in a database in the structures as detailed in Section \ref{sec:data_abstraction}.
The Analysis Pipeline module\yy{, a backend implemented in Flask,} integrates mobility-based community detection (Section \ref{sec:community_detection}), segregation measures (Section \ref{sec:segregation_measures}), and a deep learning model (Section \ref{sec:model}) enhanced with explainable AI techniques that pinpoint influential features (Section \ref{sec:model_shap}) and evaluate the impact of user-defined urban planning interventions (Section \ref{sec:model_what_if}).
Importantly, the pipeline is designed to be flexible and model-agnostic. Both the model and explainable AI techniques can be easily updated or replaced, ensuring compatibility with newer models or methods in the domain.
Finally, the Visualization module\yy{, a frontend application using React.js and D3.js,} consists of four coordinated views (Section \ref{sec:05_visual_design}), allowing users to explore analysis results and experiment with hypothetical interventions.

\section{Data Analysis Pipeline}
\label{sec:04_Data_Analysis}
This section outlines the data abstraction and the components of our analysis pipeline, which are designed to analyze segregation patterns and model mobility data effectively.

\subsection{Data Abstraction}
\label{sec:data_abstraction}
\yy{For a city of interest, we collect three complementary datasets that are commonly employed in the analysis of segregation in activity spaces~\cite{cagney2020urban}: Mobility, Sociodemographic, and Points of Interest (POI) data.}
We use the Census Block Group (CBG) as the basic spatial unit, a detailed and frequently analyzed geographic division in U.S. census data that typically contains 600-3000 residents. 
This choice of unit ensures fine-grained analysis, but our system is adaptable to other spatial divisions, such as England's Lower Layer Super Output Areas (LSOA). 
Below is a description of each dataset:

\begin{itemize}[topsep=0pt, itemsep=0.2pt, leftmargin=*]
    \item \textit{Mobility Dataset} contains the aggregated movement of individuals between various CBGs over one year. 
    It is structured as a directed weighted graph $ G = (V, E) $, where each node $ v \in V $ is a CBG, and each edge $ (v_i, v_j) \in E $ represents the flow of individuals from CBG $ v_i $ to $ v_j $, weighted by the flow volume, $ w(v_i, v_j) $.
    In our study, we utilized data provided by $E_A$, aggregated from SafeGraph~\cite{safegraph}. 
    This dataset encompasses large-scale monthly mobility flows from origin CBGs to destination places, collected from anonymous mobile devices, covering approximately 10\% of the entire U.S. population.
    
    \item \textit{Sociodemographic Dataset} contains population characteristics of each CBG (\eg, income and race).
    For an attribute $ D $ (\eg, income level), it has a matrix $ S_D = [s_{id}] $, where $ s_{id} $ is the population count of a social group $ d \in D $ (\eg, \$ 50K-100K income group) of CBG $i$.
    We obtained the data from the 5-year American Community Survey~\cite{us2020understanding} and L2 Political Academic Voter File~\cite{l2}.
    
    \item \textit{POI Dataset} contains the density (number per square kilometer) of each type of POI in each CBG. 
    It is formatted as a matrix $ P = [p_{ip}] $, where $ p_{ip} $ represents the density of POI type $ p $ in CBG $ i $.
    In our study, this dataset is also sourced from SafeGraph, containing 12 major POI types such as food, shopping, work, health, and others.
\end{itemize}

\begin{figure}
 \centering 
 \vspace{-0.2cm}
 \includegraphics[width=\linewidth]{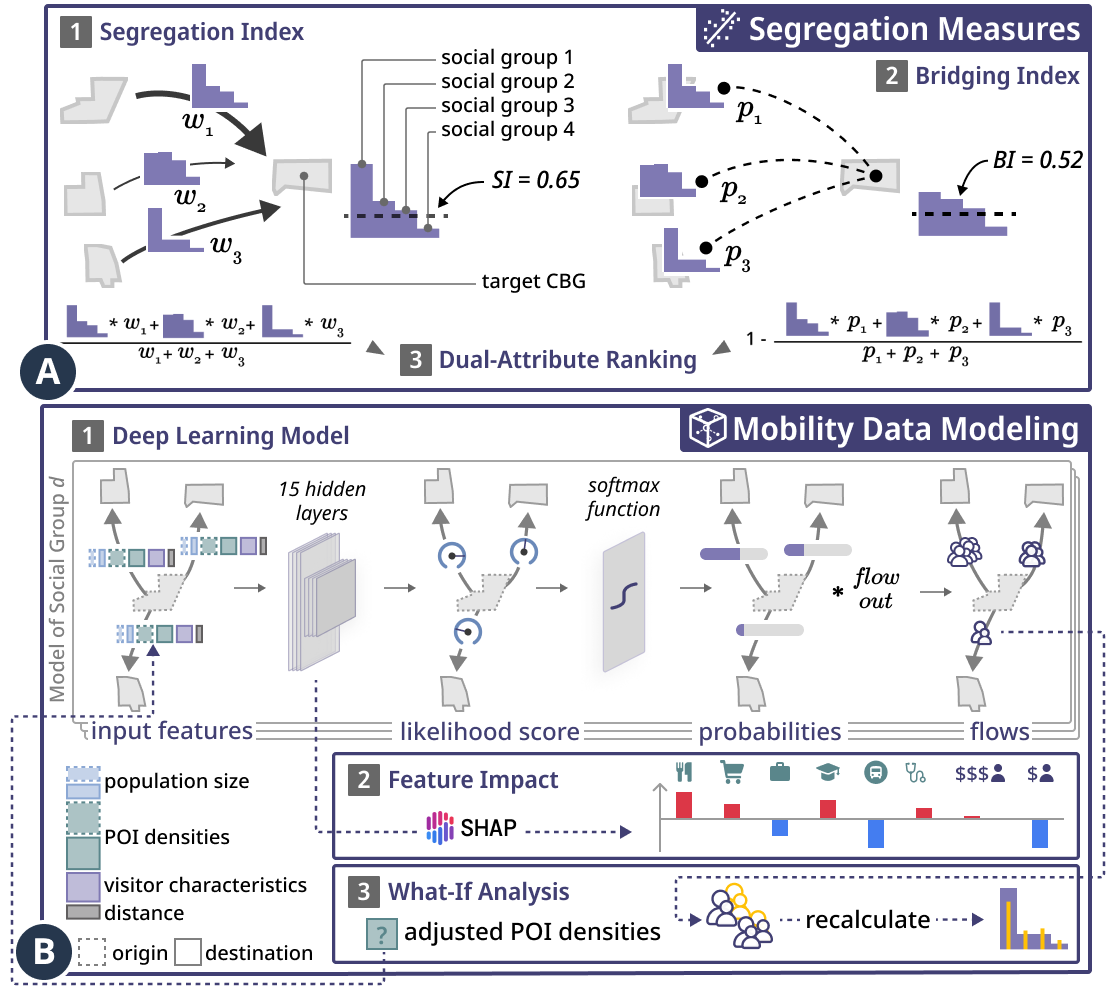}
 \vspace{-0.5cm}
 \caption{
    In the data analysis pipeline, the Segregation Measures (A) computes each CBG's Segregation Index (1) and Bridging Index (2), and then ranks them using Dual-Attribute Ranking (3).
    The Mobility Data Modeling (B) trains a Deep Learning Model for a social group (1), enhanced with Feature Impact (2) and What-If Analysis (3) modules.
 }
 \vspace{-0.6cm}
 \label{fig:data_analysis}
\end{figure}

\subsection{Mobility-based Community Detection}
\label{sec:community_detection}

In segregation studies, clustering spatial units in a city into communities with high internal movement but limited external interactions can effectively reveal data-driven social boundaries, offering a clear overview of segregated areas~\cite{huang2022unfolding, burdick2020socially} (\TIpill).

For community detection in directed weighted graphs that align with our \textit{Mobility Dataset}, multiple community detection algorithms are available, including flow-based InfoMap~\cite{rosvall2009map} and modularity-based Leiden~\cite{traag2019louvain}.
These algorithms offer distinct advantages: the Leiden algorithm tends to segment larger community structures, while the InfoMap algorithm tends to identify sub-communities~\cite{gibbs2021detecting}. 
Considering our objective of maintaining a manageable number of communities for an interpretable overview as the initial analysis, we adopt the Leiden algorithm.
The algorithm takes an edge list, where each entry contains the origin CBG $v_i$, the destination CBG $v_j$, and the flow weight $w(v_i, v_j)$, as input and iteratively optimizes modularity by reassigning CBGs to communities.
To focus on significant mobility patterns, we apply a customizable flow threshold $w_{min}$ to retain only edges where $w(v_i, v_j) \geq w_{min}$.
For interpretability, we focus on the 10 largest communities, grouping the remaining CBGs into an ``others'' community.

\subsection{Segregation Measures}
\label{sec:segregation_measures}
We employ the Segregation Index and the Bridging Index to quantify segregation patterns and intervention potential. We also implement the Dual-Attribute Ranking method to prioritize CBGs for intervention.

\subsubsection{Segregation Index}
\label{sec:segregation_index}
The segregation index (Fig.~\ref{fig:data_analysis}-A-1) is a common practice to quantify the segregation level of visitors in an area (\TIIpill).
Different formulas have been developed for the segregation index using mobility data as discussed in Section \ref{sec:related_segregation}.
In this study, we adopt the distribution unevenness recommended by our expert due to its ease of interpretation~\cite{zhang2023human}.

For any target CBG $v_{target}$ visited by $N$ origin CBGs, to calculate its segregation index of a sociodemographic attribute $D$, we first determine the proportion of residents in each social group $d$ in each origin CBG $v_i$ as $\theta_{id}$. 
Next, for each social group $d$, we calculate the proportion of visitors from $d$, $\pi_d$, defined as
\begin{equation}
\label{equation:segregation_prop}
    \pi_d(v_{target}) = \sum_{i=1}^{N} \frac{\theta_{id} \times w(v_i, v_{target})}{\sum_{i=1}^{N} w(v_i, v_{target})}.
\end{equation}
Then, the segregation index ($SI_D$), normalized between 0 and 1 by the factor $\frac{n}{2n-2}$ (higher values indicate greater segregation), is defined as
\begin{equation}
\label{equation:segregation_index}
    SI_D(v_{target}) = \frac{n}{2n-2} \times \sum_{d \in D} \left| \pi_d(v_{target}) - \frac{1}{n} \right|.
\end{equation}

\subsubsection{Bridging Index}
While traditional segregation indices highlight divisions, the bridging index (Fig.~\ref{fig:data_analysis}-A-2), proposed in~\cite{nilforoshan2023human} emphasizes opportunities for integration and connection.
It can identify possible ``bridges'' for social integration by computing the diversity of social groups interacting in a hub, assuming everyone visits their nearest hub.
However, due to limitations in our dataset, we adapt this concept by treating a CBG as a hub and focusing on the $k$ nearest CBGs to each $v_{target}$, where $k$ is adjustable. 
We first estimate the proportion of residents from each group $d$ within these CBGs who might visit $v_{target}$ as
\begin{equation}
    \pi_d(v_{target})\prime = \sum_{i=1}^{k} \frac{\theta_{id} \times p(v_i)}{\sum_{i=1}^{k} p(v_i)},
\end{equation}
where $p(v_i)$ is the population size of a origin CBG $v_i$. Then, we inverse the segregation index to formulate the bridging index ($BI_D$) as
\begin{equation}
\label{equation:bridging_index}
    BI_D(v_{target}) = 1 - \frac{n}{2n-2} \times \sum_{d \in D} \left| \pi_d(v_{target})\prime - \frac{1}{n} \right|.
\end{equation}

\begin{figure*} [t!]
 \centering 
 \vspace{-0.2cm}
 \includegraphics[width=\linewidth]{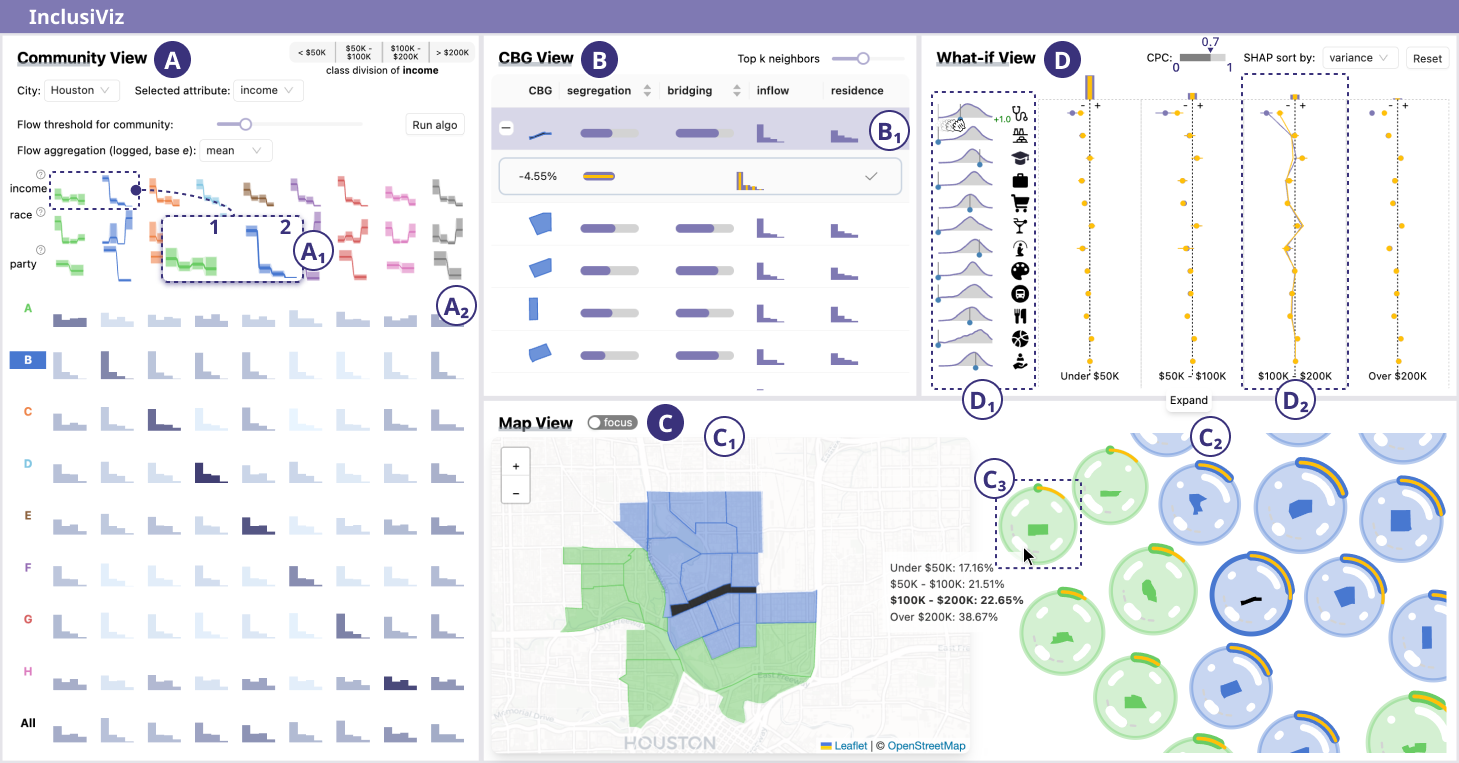}
 \vspace{-0.5cm}
 \caption{
     The interface of \systemname guides experts through a segregation analysis and iterative intervention design workflow. 
     In the \commview (A), experts analyze the sociodemographic profiles and interconnections of mobility-based communities, identifying areas with potential segregation.
     Experts then drill down into the \cbgview (B) to select a target CBG for intervention, investigating how its features influence inflow patterns from different social groups.
     Meanwhile, the \mapview (C) offers geographic context for these mobility patterns.
     Finally, experts use the \whatifview (D) to simulate urban planning interventions, with the predicted impact of those changes reflected directly in the \cbgview (B) and the \mapview (C).
 }
 \vspace{-0.4cm}
 \label{fig:system-visual}
\end{figure*}

\subsubsection{Dual-Attribute Ranking}
\label{sec:TOPSIS_ranking}
Research by Nilforoshan \et~\cite{nilforoshan2023human} highlights a strong inverse relationship between bridging and segregation indices.
While relocating residents to bridge different social groups is beneficial, it is too ethically complex and thus impractical for our study~\cite{sanbonmatsu2011moving}. 
In consultation with our expert, we shifted focus towards CBGs with high bridging index (diverse neighborhoods) yet still experiencing high segregation.
These offer the potential for interventions to increase visitor diversity (\TIIIpill).

A Multi-Criteria Decision Making (MCDM) method can help us more easily prioritize the CBGs with both high segregation index $SI$ and bridging index $BI$.
Among various MCDM techniques, we selected the Technique for Order of Preference by Similarity to the Ideal Solution (TOPSIS)~\cite{lai1994topsis} for its simplicity and effectiveness.
TOPSIS ranks CBGs based on their proximity to both ideal (high $BI$ and $SI$) and negative-ideal (low $BI$ and $SI$) benchmarks (Fig.~\ref{fig:data_analysis}-A-3). 
It effectively surfaces CBGs that not only are positioned well for fostering integration but also currently suffer from high segregation.

\subsection{Mobility Data Modeling}

\subsubsection{Deep Learning Model}
\label{sec:model}
We developed a deep learning model adapted from Deep Gravity~\cite{simini2021deep} (Fig.~\ref{fig:data_analysis}-B-1), the state-of-the-art deep learning framework designed to predict mobility flows based on environmental features including distance, population size, land use, and densities of different types of POI.
We made two key adaptations for segregation analysis, which, as evaluated in Section~\ref{sec:model_evaluation}, improved the model's prediction precision.

\begin{itemize}
    \item Social group segmentation: 
        We divide the \textit{Mobility Dataset} ($G$) into segments ($G_d$) for each social group $d \in D$, enabling separate models to capture group-specific mobility patterns.
    \item Visitor characteristics inclusion:
        Inspired by Moro \et~\cite{moro2021mobility} that models the preference of people to visit the places where they are the majority in visitors, we include the sociodemographic characteristics of visitors computed using Equation \ref{equation:segregation_prop} as part of the input vectors. 
        This aims to model how different social groups respond to the visitor characteristics of a destination area.
\end{itemize}

Our adapted model's input for an origin-destination CBG pair $(v_i, v_j)$ is $concat[p_i, p_j, x_i, x_j, \pi_j, dis_{i,j}]$, where $p_i$, $p_j$ are the population sizes, $x_i$, $x_j$ are densities of different types of POIs in the origin and destination, $\pi_j$ is a vector for the visitor characteristics of the destination, and $dis_{i,j}$ is their geographical distance. 
We employ a neural network architecture consistent with the original Deep Gravity model with 15 hidden layers, combining wider layers (256 neurons) and narrower layers (128 neurons) with LeakyReLU activations.
The network predicts a mobility likelihood score $s(v_i, v_j)$ for each origin-destination CBG pair. For each origin CBG $v_i$, it calculates these scores for $k$ randomly selected destinations, and here $k$ is set to be 30 to speed up the computation.
The likelihood scores are transformed into predicted probabilities using the softmax function, and the predicted flow volume from $v_i$ to $v_j$, $\hat{w}(v_i, v_j)$, is calculated by multiplying the total number of individuals from $v_i$ to its $k$ destinations with its probability, formulated as
\begin{equation}
    \hat{w}(v_i, v_j) = \left( \sum_{l=1}^{k} w(v_i, v_l) \right) \cdot \frac{e^{s(v_i, v_j)}}{\sum_{l=1}^{k} e^{s(v_i, v_l)}}.
\end{equation}

\subsubsection{Feature Impact}
\label{sec:model_shap}
To understand how individual features influence predictions in our deep learning model (\TIVpill), we utilize SHAP values~\cite{lundberg2017unified} (Fig.~\ref{fig:data_analysis}-B-2), a widely adopted explainable AI technique.
SHAP quantifies each feature's contribution to the predicted likelihood of mobility flows, capturing both the direction and magnitude of its influence.
Unlike Partial Dependence Plots (PDP), which show only average effects, SHAP provides more granular, instance-specific insights. 
We chose SHAP over other local explanation techniques like LIME~\cite{ribeiro2016should} for its basis in Shapley values, ensuring fair and consistent feature attribution across instances for more stable explanations.

We use \texttt{DeepExplainer} module from the SHAP library, which is well-suited for approximating feature contributions within deep neural networks.
For each social group's model, SHAP values of the features are computed for the predicted flow $\hat{w}(v_i, v_{target})$ between a pair of origin and target CBG. 
For efficiency, we precompute 50 centroids via K-means clustering as the background dataset for the SHAP explainer, ensuring representative feature impact approximations.

\subsubsection{What-If Analysis}
\label{sec:model_what_if}
To allow users to explore the potential impact of hypothetical urban planning changes on mobility patterns (\TVpill), we developed a What-If Analysis module to enhance the deep learning model (Fig.~\ref{fig:data_analysis}-B-3).
When some POI-related features in the target CBG ($v_{target}$) are modified as $x_{target}\prime$, the function modifies the input vector for each pair of origin and target CBG $(v_i, v_{target})$, resulting in a new set of input vectors $concat[p_i, p_{target}, x_i, x_{target}\prime, \pi_{target}, dis_{i, {target}}]$.
These modified vectors are then fed to our trained models to predict new mobility flows \(\hat{w}_{after}(v_i, v_{target})\) from different social groups.
To quantify the impact of the change, we calculate the difference between the predicted flows before and after the hypothetical intervention, defined as
\begin{equation}
    \Delta \hat{w}(v_i, v_{target}) = \hat{w}_{after}(v_i, v_{target}) - \hat{w}_{before}(v_i, v_{target}),
\end{equation}
and increment the predicted differences to actual flow volumes accordingly as the new flow volumes result from feature adjustments.
\section{Visual Design}
\label{sec:05_visual_design}

\systemname consists of four views to fulfill the analysis goals introduced in Section~\ref{sec:03_Background_VisualizationDesignTasks}: \commview (Fig.~\ref{fig:system-visual}-A), \cbgview (Fig.~\ref{fig:system-visual}-B), \mapview (Fig.~\ref{fig:system-visual}-C), and \whatifview (Fig.~\ref{fig:system-visual}-D).
We demonstrate the seamless analysis flow using the four views with an illustrative use case derived:
\yy{A social scientist investigates income segregation in a city, examining contributing factors and exploring potential interventions to mitigate observed segregation.}
After analyzing the interconnection between detected mobility-based communities (\TIpill) and comparing their sociodemographic distributions (\TIIpill) in the \commview, he notices one with a significant low-income concentration, indicating segregation.
Drilling down into this community in the \cbgview, he identifies a target CBG with high segregation and potential for intervention (\TIIIpill).
Cross-referencing inflow patterns of the target CBG in the \mapview with the feature impact in the \whatifview, he observes a lack of health-related POIs in the target CBG, which could explain its limited appeal to higher-income groups (\TIVpill).
Finally, using the \whatifview to simulate an increase in such POIs, he receives feedback suggesting this intervention could reduce segregation (\TVpill).

\subsection{Community View}
The \commview (Fig.~\ref{fig:system-visual}-A) provides an overview for exploring mobility-based community structures within the city and their sociodemographic characteristics.
It allows experts to customize the city and attribute of interest and the flow threshold, and detect communities using the Leiden algorithm as described in Section~\ref{sec:community_detection}.
Then, it supports the comparison of community sociodemographic profiles using the \commsig (Fig.~\ref{fig:system-visual}-A1) (\TIIpill) and inter-community flows in the \flowmat (Fig.~\ref{fig:system-visual}-A2) (\TIpill).
Finally, experts can indicate a community of interest for the following analysis.

\begin{figure} [t]
 \centering 
 \vspace{-0.2cm}
 \includegraphics[width=\linewidth]{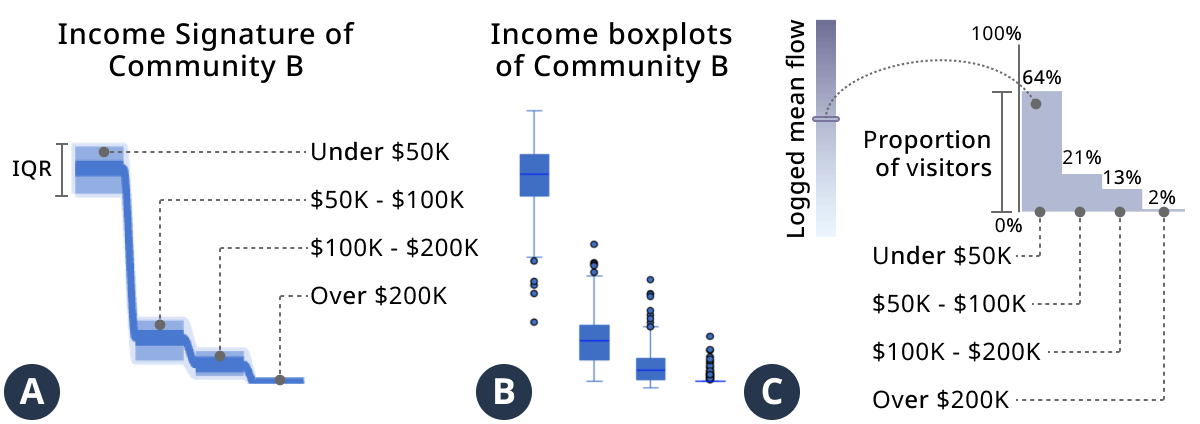}
 \vspace{-0.5cm}
 \caption{
     (A) The income \commsig design to show the income distributions of CBGs in community B. 
     (B) shows an alternative design: traditional boxplots.
     (C) The visual encoding of a cell in the \flowmat.
 }
 \vspace{-0.6cm}
 \label{fig:signature}
\end{figure}

\textbf{\commsig.}
To support the efficient analysis of the sociodemographic profiles of detected communities, a \commsig of each attribute accompanies each community, displayed in its corresponding color (\TIIpill).
The top row features the signatures of the chosen attribute, while subsequent rows provide additional attribute signatures as auxiliary information.
A signature (Fig.~\ref{fig:signature}-A) resembles horizontally aligned boxplots (Fig.~\ref{fig:signature}-B), with only the interquartile range (IQR) preserved and linked by smooth lines.
Within each IQR box in a signature, its height represents the range of proportions for a social group within the central 50\% of CBGs in the community, with the median line highlighted in a darker color for emphasis.
A detailed legend about the signatures of an attribute will be shown upon hovering over the question mark icon beside the attribute name.

\textit{Justification:}
We initially displayed distributions using traditional boxplots (Fig.~\ref{fig:signature}-B). 
However, our domain expert found them visually overwhelming when comparing sociodemographic profiles across multiple communities. 
He suggested that at the early analysis stage, social scientists often focus on the distributions within the majority of CBGs in a community to gain an overall impression. 
Therefore, we designed the \commsig to emphasize the essential IQR information, representing the central concentration of CBGs, while simplifying the visual presentation.
Connecting IQRs with lines enhances visual continuity, aiding efficient comparison.
Our expert further recognized that this design streamlined his analysis, allowing him to rapidly identify similarities and differences between community attribute distributions based on the signature shapes.

\textbf{\flowmat.}
The \flowmat visualizes inter-community mobility patterns within the city (\TIpill), which has been segmented into N communities.
This $(N+1) \times (N+1)$ matrix resembles the adjacency matrix of a directed weighted graph. 
In the matrix, rows represent community outflows, columns represent inflows, and the final column/row depicts total flows.
Each matrix cell (Fig.~\ref{fig:signature}-C) encodes the visitor flow between two communities, and bars within the cell represent the proportions of different social groups (consistent with the encoding in the \commsig).
A log-normalized scale represents flow volume through the background color of the bars, and the flow volume can be aggregated as mean, median, sum, and standard deviation via the ``Flow aggregation'' selector above. 
Community labels on the left act as selectors for experts to choose a community of interest.

\textit{Justification:}
While a node-link visualization was considered, we adopted the matrix representation to visualize the multivariate flows and their attributes for its effectiveness of matrices for comparing edge attributes~\cite{nobre2019state, alper2013weighted}.
In our case, these attributes represent the crucial distribution of social groups within the flows between communities.

\subsection{CBG View}
\label{vis:cbg_view}
The \cbgview (Fig.~\ref{fig:system-visual}-B) presents a \cbgTable to identify CBGs with intervention potential for mitigating segregation within a chosen community (\TIIIpill).

The interactive \cbgTable inspired by the Ranking View in SmartAdp~\cite{liu2016smartadp} displays a list of CBGs ranked by TOPSIS, as described in Section~\ref{sec:TOPSIS_ranking}, that prioritizes the CBGs in diverse neighborhoods but experiencing high segregation (\TIIIpill).
Each row showcases a CBG with the geographical boundary shape, segregation index $SI$ and bridging index $BI$ (with bars representing their magnitudes), and proportions of social groups behind these two indices (encoded with bar charts consistent with the \flowmat).
Experts can adjust the $k$ neighbors parameter at the top to modify the bridging index calculation and reorder the table.
Selecting a CBG row will set it as the target CBG, allowing detailed feature impact exploration in the \whatifview and inflow pattern analysis in the \mapview.

\subsection{Map View}
\label{vis:map_view}
The \mapview (Fig.~\ref{fig:system-visual}-C) provides geographical information, juxtaposing the \realmap (Fig.~\ref{fig:system-visual}-C1) for geographical context for the detected communities (\TIpill) and the \dorlingmap (Fig.~\ref{fig:system-visual}-C2) to analyze the detailed flows between CBGs (\TIVpill).

\textbf{\realmap}. It displays the city's map with color-coding CBGs according to their community assignment. It offers two modes: ``full mode'' reveals the entire city, while ``focus mode'' highlights a selected target CBG (in black) and its closest $k$ neighbors, where $k$ adjusts based on settings in the \cbgTable of the \cbgview (Section~\ref{vis:cbg_view}).

\textbf{\dorlingmap}. It employs a Dorling layout~\cite{dorling1996area} \yy{over traditional choropleth maps to better support glyph-based multivariate encoding of sociodemographic and flow patterns while preserving relative CBG positions}.
Specifically, a \bubbleglyph (Fig.~\ref{fig:system-visual}-C3) visualizes the detailed information about a CBG.

\begin{figure} [t]
 \centering 
 \vspace{-0.2cm}
 \includegraphics[width=\linewidth]{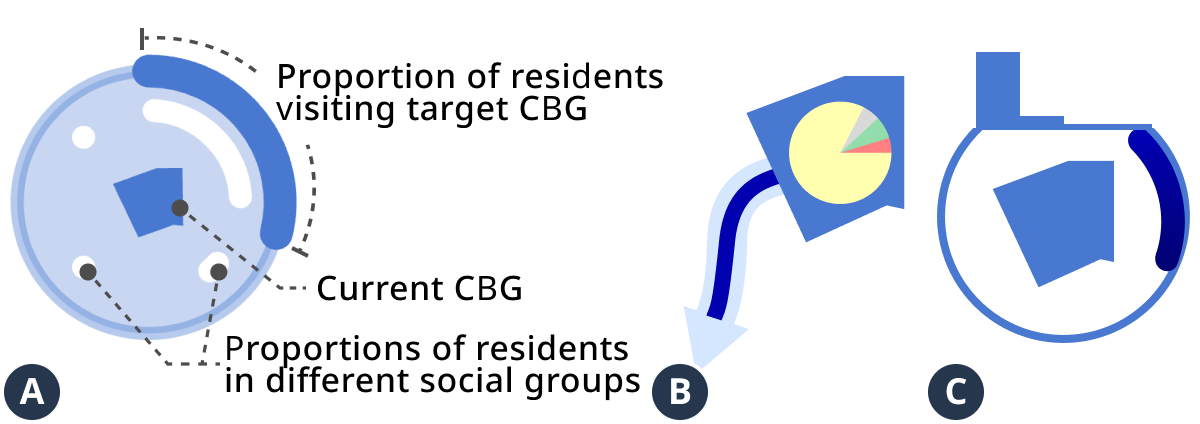}
 \vspace{-0.5cm}
 \caption{
     (A) The \bubbleglyph design to demonstrate one CBG's resident sociodemographics and flow patterns to target CBG.
     (B) and (C) show two alternative glyph designs.
 }
 \vspace{-0.6cm}
 \label{fig:glyph}
\end{figure}

\textbf{\bubbleglyph.}
Inspired by the concept of social filter bubbles that isolate people into homogeneous individuals, we designed the metaphor-based \bubbleglyph to showcase one CBG's resident demographics and flow patterns, as shown in Fig.~\ref{fig:glyph}.
The background color of the glyph matches the CBG's assigned community, and the glyph's \yy{area} represents the population size of the CBG.
The CBG's actual geographical boundary shape is at the center inside the glyph to aid in connecting with \realmap and \cbgTable.
A circle of white proportion arcs around the shape represents the proportions of different social groups in the CBG's resident population.
The arcs resemble the reflection light on a bubble: the dominance of one arc will create a bubble-like appearance, visually indicating segregation.
Hovering over an arc highlights its full extent and reveals detailed group information in a tooltip.
A thicker flow arc around the boundary represents the proportion of residents visiting the target CBG.

\textit{Justification:}
We designed the bubble glyph after carefully considering two alternatives with our expert (Fig.~\ref{fig:glyph}-B, C).
Design B encodes the proportion of outgoing visitors from each CBG to the target CBG by the thickness of a darker inner edge, and represents the proportions of different social groups within that CBG with a pie chart.
However, as the number of glyphs increases on the map, these edges begin to cross over each other, leading to severe visual clutter.
Design C places social group proportions as a bar chart consistent with \flowmat, which saves space.
However, this layout limits the flow arc to less than a full circle, which is counter-intuitive and potentially misleading for chart reading.
In contrast, the \bubbleglyph remains the expert's favored design as it is intuitive and aesthetically pleasing.
Its metaphor-based approach vividly communicates segregation while avoiding issues like edge crossings, and it facilitates clearer comparisons of demographic proportions across multiple CBGs.

\subsection{What-If View}
\label{vis:whatif_view}
The \whatifview (Fig.~\ref{fig:system-visual}-D) enables users to explore how environmental features influence the number of visitors from different social groups to a target CBG (\TIVpill) and design hypothetical interventions for reducing segregation for the CBG (\TVpill).
At the top, a CPC bar (introduced in Section~\ref{sec:model_evaluation}) shows the trained model's predictive accuracy in the target CBG.
Below, it juxtaposes a \featControl (Fig.~\ref{fig:system-visual}-D1) and \featImpacts for different social groups (Fig.~\ref{fig:system-visual}-D2).

\textbf{\featControl}. A list of Kernel Density Estimate (KDE) plots depicts the global distributions of features across the entire dataset, with the current values for the target CBG indicated by draggable lines. Users can interactively adjust feature values by dragging these lines horizontally (\TVpill). 
Expert feedback indicates that mitigating segregation is more feasible through adjustments in POI rather than demographic characteristics, so the panel lists only POI-related features by default, with an option to expand the panel to access the full feature list, including both sociodemographic and POI-related factors.

\textbf{\featImpact}. A \featImpact illustrates how features in the target CBG influence the flow of visitors from a social group in neighboring CBGs (\TIVpill). At the top of each column, the proportion of visitors from the social group is visualized as a bar chart, consistent with the chart in the \cbgview. Below, each feature's average SHAP value is shown as a dot with error bars for standard deviation, indicating the strength of its influence on attracting or repelling visitors from the social group to the target CBG.
By default, features are ordered by the sum of their SHAP value magnitudes. 
Following expert recommendations, the system also allows sorting by the variance in SHAP values across social groups, facilitating inter-group comparisons of mobility behaviors.
For more detailed instance-level explanations, when users hover over a social group arc in the \bubbleglyph of a specific CBG $v_i$, a line connecting the individual SHAP values overlays the dots, illustrating how features in the target CBG $v_{target}$ affect visitor behavior from that social group within $v_i$.

\subsection{Providing Real-time Feedback}
\label{vis:interaction}
To enable users to develop and evaluate urban planning interventions with real-time feedback (\TVpill), adjustments made to the target CBG’s features in the \featControl of the \whatifview are dynamically fed into two modules: the deep learning model’s Feature Impact (Section~\ref{sec:model_shap}) and What-If Analysis (Section~\ref{sec:model_what_if}). These modules recalculate the SHAP values for the features and predict new mobility flows from different social groups to the target CBG. Additionally, a new Segregation Index (Section~\ref{sec:segregation_index}) is computed for the target CBG.

This process triggers real-time updates across the visualization components, providing users with immediate feedback for intervention evaluation and refinement. The updates are as follows:

\cbgTable (Section~\ref{vis:cbg_view}): The corresponding row of the target CBG will become expandable, with sub-rows representing intervention strategies (Fig.~\ref{fig:system-visual}-B1).
Predicted changes to the segregation index and inflow proportions of social groups are visualized with thin gold bars overlaying the original bars. 
Users can save, edit, and delete strategies using the icons on the right side. 

\dorlingmap (Section~\ref{vis:map_view}): In glyphs, thinner gold flow arcs will overlay the original flow arcs to visualize flow changes resulting from interventions (Fig.~\ref{fig:system-visual}-C3).

\featImpact (Section~\ref{vis:whatif_view}): Recalculated average SHAP values for the features are displayed as gold dots alongside the original purple dots for comparison (Fig.~\ref{fig:system-visual}-D2). A gold line connecting the recalculated instance-level SHAP values (if a social group in a CBG is hovered) also appears beside the original purple line.
\section{Evaluation}
\label{sec:evaluation}
In this section, we present a quantitative analysis of our deep learning model, followed by two case studies conducted with experts in segregation analysis. 
\yy{We introduced the system’s workflow and visual encodings to the experts and collaborated with them to develop the cases.}
For case 1, we invited $E_B$, a marketing data scientist with three years of experience in income-based segmentation, to income segregation.
For case 2, we worked with our domain expert $E_A$, introduced in Section~\ref{prob_formulation}, to explore partisan segregation.
Besides, we invited two more experts ($E_C$ and $E_D$) for the expert interviews detailed in Section~\ref{design_implication}.
$E_C$ has four years of experience as a geospatial analyst, focusing on agent-based modeling (ABM) of human mobility.
$E_D$ is a doctoral student researching segregation on social media for two years.

\subsection{Quantitative Experiments}
\label{sec:model_evaluation}

\parasum{Performance.}
To assess our adaptations to the original Deep Gravity model (social group segmentation and visitor characteristics inclusion as described in Section \ref{sec:model}), we tested two baseline settings and three enhanced settings. 
The two baselines are the original Gravity (G) and Deep Gravity (DG) models. The three enhanced settings include DG with social group segmentation (DG+S), DG with visitor characteristics (DG+V), and the combination of both (DG+S+V), which is our final adopted version.
Consistent with case 1, we used Houston data, segmented social groups by income, and included income and race as visitor characteristics.
We assessed model performance using \yy{five widely recognized} metrics: Common Part of Commuters (CPC) measures the similarity between predicted and target flow volumes; 
Jensen-Shannon Divergence (JSD) quantifies the difference between their distributions; 
Pearson Correlation assesses the strength of the linear relationship;
Root Mean Squared Error (RMSE) and \yy{Normalized Root Mean Squared Error (NRMSE)} quantify the average prediction error\yy{, with NRMSE offering interpretability by normalizing errors relative to the target flow range}. 
Higher CPC and Pearson values, and lower JSD, RMSE, and NRMSE indicate better performance.
Table \ref{tab:algo-results} shows that DG+S+V achieved the best results across all metrics.
\footnote{We did a comprehensive evaluation of different cities and population densities. Please refer to the appendix for more information.}


\begin{table}[h!]

\centering
\small
\vspace{-0.2cm}
\begin{tabular}{l|c|c|c|c|c}
\toprule
Model & CPC (↑) & JSD (↓) & Pearson (↑) & RMSE (↓) & \yy{NRMSE (↓)} \\
\midrule
    DG+S+V & \textbf{0.6197} & \textbf{0.3497} & \textbf{0.7733} & \textbf{200.0248} & \yy{\textbf{0.0069}} \\
    DG+V & 0.6100 & 0.3580 & 0.6775 & 218.7619 & \yy{0.0076}\\
    DG+S & 0.6125 & 0.3565 & 0.7362 & 210.4794 & \yy{0.0072}\\
    DG & 0.5979 & 0.3684 & 0.6125 & 230.5540 & \yy{0.0078}\\
    G & 0.4605 & 0.4517 & 0.3862 & 245.4852 & \yy{0.0083}\\
\bottomrule
\end{tabular}
\vspace{-0.2 cm}
\caption{Performance comparison across different model settings}
\vspace{-0.5 cm}
\label{tab:algo-results}

\end{table}

\parasum{Time complexity.}
We performed model training and inference on a 16GB MacBook Pro (Apple M1, no GPU). 
For Houston (around 1500 CBGs), in our case, individual model training took approximately 2 minutes. 
Therefore, we pre-trained models with expert-identified features offline. 
Model inference during what-if analysis (Section \ref{sec:model_what_if}) was fast (typically within 2 seconds), ensuring real-time feedback.

\begin{figure*} [t!]
 \centering 
 \vspace{-0.5cm}
 \includegraphics[width=\linewidth]{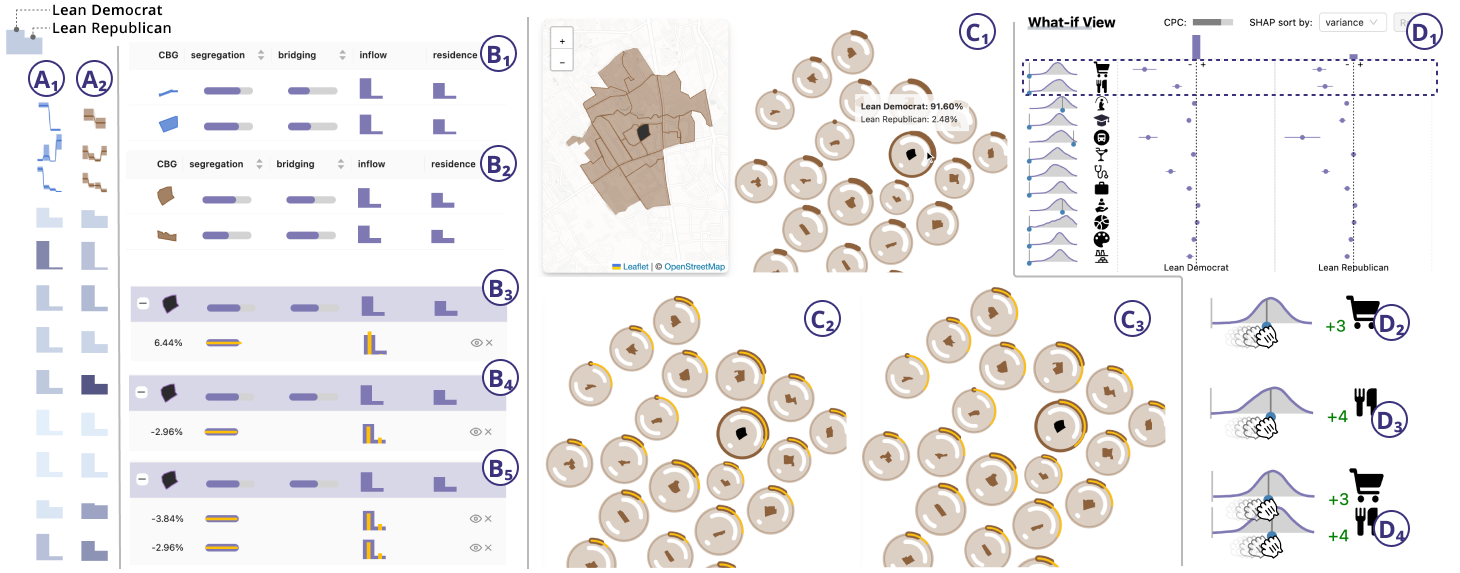}
 \vspace{-0.5cm}
 \caption{
     Case 2 - \systemname guided expert $E_A$ in analyzing partisan segregation and designing interventions. 
     He first compared the sociodemographics of communities (A1, A2) and searched for promising CBGs with intervention potential (B1, B2).
     He then analyzed the partisan preferences in mobility to the target CBG (C1, D1).
     He tried increasing food-related POIs (D2) to attract diverse visitors but it surprisingly worsened segregation (B3, C2).
     In contrast, increasing shopping facilities (D3) reduced segregation (B4).
     Finally, a combined intervention (D4) proved most effective (B5, C3).
 }
 \vspace{-0.6cm}
 \label{fig:system-case}
\end{figure*}

\definecolor{commAColor}{HTML}{51c44a} 
\definecolor{commBColor}{HTML}{4878d0}
\definecolor{commEColor}{HTML}{8c613c}
\newcommand{\commA}{Community \textcolor{commAColor}{\textbf{A}}\xspace}
\newcommand{\commB}{Community \textcolor{commBColor}{\textbf{B}}\xspace}
\newcommand{\commE}{Community \textcolor{commEColor}{\textbf{E}}\xspace}

\subsection{Case Studies}
Our cases focus on understanding and mitigating urban segregation in Houston City, the fourth-most populous city in the United States, known for its income inequalities~\cite{houstonincome} and partisan polarization~\cite{houstonpartisan}.

\subsubsection{Case1: A Prescription for Income Harmony}
We invited $E_B$, an experienced marketing data scientist working on income-based segmentation, to explore the first case about \textbf{income segregation}.
The income attribute uses the census data from 5-year American Community Survey and classifies the income into four groups from low to high: ``Under \$50K'', ``\$50K - \$100K'', ``\$100K - \$200K'', and ``Over \$200K''.
As suggested by $E_B$, the visitor characteristics in the input feature include both race and income.

\textbf{Overview of income disparities.} 
$E_B$ compared income signatures (\TIIpill) in the \commview (Fig. \ref{fig:system-visual}-A), and \commB (blue) emerged with a distinct segregation profile, characterized by a significant first box representing the low-income population (Fig. \ref{fig:system-visual}-A1-2).
In contrast, \commA (green), adjacent to Community B, exhibits greater economic diversity, with its signature showing a more even distribution across groups (Fig. \ref{fig:system-visual}-A1-1).
The \realmap (Fig. \ref{fig:system-visual}-C1) helped him localize them: \textit{``Community B is north of downtown, where neighborhoods face high poverty; Community A is located southwest of downtown, with booming areas like Sugar Land.''}

He then turned to the \flowmat (Fig. \ref{fig:system-visual}-A2) to understand inter-community connections (\TIpill), finding that Community B displays the most extreme income segregation in both incoming (last row) and outgoing (last column) mobility patterns.
Additionally, the lighter background shading of Community A's flow bar chart indicates minimal interaction with Community B.
\textit{``Despite their proximity, these communities rarely interact, deepening the segregation in activity spaces. How can we bridge this divide?''} $E_B$ wondered.
Therefore, he clicked Community B's label in the left column to focus on Community B.

\textbf{Targeting a CBG for intervention.}
Targeting specific CBGs within Community B for interventions (\TIIIpill), $E_B$ utilized the \cbgTable (Fig. \ref{fig:system-visual}-B), focusing on those with high segregation and bridging indices.
Selecting the first CBG as the target triggered a series of visual updates: it was highlighted in black on both the \cbgTable and \realmap.
The \dorlingmap (Fig. \ref{fig:system-visual}-C2) and \featImpacts (Fig. \ref{fig:system-visual}-D2) also updated, displaying the target CBG's neighborhood connections and its feature impact.

\textbf{Analyzing inflow patterns and revealing intervention needs.}
$E_B$ further delved into inflow patterns to the target CBG, aiming to uncover the reasons behind its segregation (\TIVpill).
The \dorlingmap's \bubbleglyph (Fig. \ref{fig:system-visual}-C2) revealed a clear trend: the target CBG and nearby CBGs in Community B were dominated by the lowest income group (longest white arc). In contrast, five CBGs from Community A to the west had a more balanced income distribution but sent few visitors to the target CBG.

The CPC bar showed the model predicted 70\% of the Common Part of Commuters to the target CBG accurately.
To investigate the disparity, the expert compared the \featImpacts (Fig. \ref{fig:system-visual}-D2) of the four social groups. 
To explore inter-group differences, he sorted the SHAP values by variance, revealing that health-related POI density ranked at the top.
The \featControl (Fig. \ref{fig:system-visual}-D1) showed no such amenities in the target CBG, while the \featImpacts revealed its stronger negative influences on the ``\$50K - \$100K'' and ``\$100K - \$200K'' groups.
This aligned with the expert's insight: \textit{``Health-related amenities tend to attract middle-class populations, as lower-income groups face affordability barriers and higher-income individuals often use private services. This market gap reinforces segregation.''}

\textbf{Testing an intervention strategy for integration.}
Inspired by his findings, $E_B$ turned to the \featControl (Fig. \ref{fig:system-visual}-D1) to design an intervention to reduce income segregation (\TVpill).
He simulated increasing health-related POIs in the target CBG by 1 unit per square kilometer.
The predicted outcome showed a 4.55\% decrease in segregation, as reflected in the expanded row of the \cbgTable.
Additionally, the \dorlingmap visualized an increased flow from Community A to the target CBG, depicted by longer gold arcs.
Hovering over the arc for the ``\$100K - \$200K'' group in a neighboring CBG (Fig. \ref{fig:system-visual}-C3), $E_B$ observed a significant rise in the instance-level SHAP value for health POIs in the \featImpact (Fig. \ref{fig:system-visual}-D2), confirming that the intervention attracted more middle-class visitors.
\textit{``This highlights the importance of strategic placement,''} $E_B$ concluded, \textit{``Adding resources isn't enough – it's about targeting them to bridge communities. Much like expanding a product line to attract a wider customer base.''}

This case demonstrates how \systemname assists in understanding the urban segregation landscape, its possible causes, and how to design data-driven interventions. 
While tailored for social science, this example highlights the system's value for other fields, like marketing, where attracting diverse groups is a strategic goal.

\subsubsection{Case2: From Partisan Lines to Shared Spaces}
We invited $E_A$, our domain expert focusing on political polarization research, to explore the second case about \textbf{partisan segregation}.
We used the L2 Political Academic Voter File to obtain partisan membership data and  classified non-Democratic or Republican parties as either ``Lean Democrat'' or ``Lean Republican,'' excluding independent or non-partisan individuals.
Proportions of these two partisan groups were then calculated for each CBG and included in the Sociodemographic Dataset (as described in Section \ref{sec:data_abstraction}). 
Following $E_A$'s suggestion, both income and partisan memberships were used as visitor characteristics to reveal the interplay between income levels and political beliefs.

\textbf{Searching for CBGs with intervention opportunities.}
After segmenting the city into communities (\TIpill), $E_A$ navigated through each community, searching for CBGs with intervention opportunities (\TIIpill{}, \TIIIpill). 
\commB (blue), predominantly Lean Democrat, initially caught his eye (Fig. \ref{fig:system-case}-A1). 
However, a closer look revealed a lack of CBGs with both a high segregation index and a high bridging index (Fig. \ref{fig:system-case}-B1), making it less ideal for his intervention plan.
Redirecting his focus, $E_A$ discovered \commE (brown), a politically mixed area (Fig. \ref{fig:system-case}-A2) with a promising landscape, with several CBGs flagged for both high segregation and bridging potential (Fig. \ref{fig:system-case}-B2). 
Therefore, he selected Community E and chose its first CBG for in-depth analysis.

\textbf{Analyzing partisan preferences in mobility.}
$E_A$ examined inflow patterns using the \dorlingmap (Fig. \ref{fig:system-case}-C1) and \featImpacts (Fig. \ref{fig:system-case}-D1) (\TIVpill), observing a clear Democratic majority in the target CBG. Republican-leaning neighboring CBGs, especially on the west side, showed minimal inflow, indicating a political divide. 
Sorting features by SHAP value variance, $E_A$ noted that the densities of shopping and food-related facilities had the most significant differential impact on mobility patterns for both groups. 
Despite the absence of both in the target CBG, $E_A$ inferred that Lean Democrats were more negatively affected by the lack of shopping venues, whereas Lean Republicans were more deterred by the absence of food outlets.
This observation aligns with the idea of ``political consumption,'' where consumer preferences reflect political identities: Democrats may favor niche shopping venues, while Republicans might prefer familiar food outlets.

\textbf{Exploring synergies to bridge divides.}
$E_A$ used the \featControl to test suitable interventions for segregation mitigation (\TVpill).
He began by increasing the shopping facility density to 3 units per square kilometer (Fig. \ref{fig:system-case}-D2), which resulted in a 6.44\% increase in the segregation index (Fig. \ref{fig:system-case}-B3).
Examining the \dorlingmap (Fig. \ref{fig:system-case}-C2), he observed,
\textit{``While we saw an increase in visitors from Republican-dominant CBGs, there was also a surge in visitors from the nearby Democrat-dominant CBGs, which worsened the segregation.''}

Resetting the current intervention, $E_A$ increased food facility density, noting a slight reduction in segregation. Advancing to 4 units per square kilometer (Fig. \ref{fig:system-case}-D3), he witnessed a 2.96\% drop (Fig. \ref{fig:system-case}-B4), with a notable increase in Republican-leaning visitors.
\textit{``Though both food and shopping facilities attract visitors, they have unique characteristics. 
Existing literature also shows different POIs have different contributions to segregation~\cite{zhang2023human}. Places that require more face-to-face interaction, in this case, the local grocery stores, can be more segregated.''} $E_A$ commented and recorded this intervention.

Intrigued by the interaction between shopping and food POIs, he simultaneously increased both types of amenities: shopping to 3 units per square kilometer and food to 4 units per square kilometer (Fig. \ref{fig:system-case}-D4).
Interestingly, this time, the segregation index did not increase because of the food facilities, but conversely, it decreased by 3.84\% (Fig. \ref{fig:system-case}-B5).
From the updated \dorlingmap (Fig. \ref{fig:system-case}-C3), it could be seen that Republican-dominated CBGs on the west side had more significant increases in visitor proportions to the target CBG.
\textit{``I thought the change would be linear and counteracted, but I didn't expect their combined effect to further decrease the segregation,''} $E_A$ pondered.
\textit{``Previously, we always treated POIs individually, and the combined effect of POIs is rarely touched, but such effect seems to be captured by the deep learning model. 
One hypothesis is those facilities co-create shared experiences of consumption and leisure, making the target CBG an even more inclusive multi-use community center for people holding different political beliefs.''}
Finally, $E_A$ thanked the system for shedding light on his future research: \textit{``In my following study, I may focus on this interesting effect of mixed-use area in the partisan segregation.''}

In summary, this case highlights that \systemname can help experts develop data-informed and feedback-driven urban planning insights to promote political integration and help researchers gain new insights that can be hard to reveal with traditional social science workflow.
\section{Discussion}
The two case studies demonstrate the usability and effectiveness of \systemname in helping experts understand and mitigate different types of urban segregation.
To enrich our findings further, we interviewed two more experts ($E_C$ and $E_D$).
\yy{We presented the case study results and invited them to explore the system freely, summarizing their feedback.}

\subsection{Design Implications}
\label{design_implication}
We summarize insights from the feedback of four experts.

\textbf{Applications and target users.} 
All experts praised the usefulness of \systemname, suggesting its potential applications beyond the current user base.
For the \commpill unity and \cbgpill-level analysis, $E_B$ and $E_C$ found the progressive workflow efficient for geospatial analysts to explore complex mobility data in unfamiliar cities without extensive custom coding.
$E_D$ expected to utilize the \commview and \dorlingmap to visualize communities and echo chambers on social media.
For the \poipill-level Analysis, $E_A$ and $E_B$ saw it as a bridge between researchers and decision-makers, offering benefits in various scenarios. 
Besides social scientists, $E_A$ shared that social activists and NGOs promoting social well-being in communities could use the system to assess the potential impact of municipal policies: \textit{``Imagine using \systemname to present different what-if scenarios dynamically; it could influence discussions and even the final planning.''}
$E_B$ underscored the system's value for professionals in fields like marketing.
He compared it with TensorFlow's What-If tool, praising \systemname for its visually engaging outputs and intuitive comparisons that enhance presentations.
\yy{
Although experts found \systemname easy to use after our introduction, they suggested an onboarding tutorial with background context could further lower the learning curve for novices navigating different views.
}

\textbf{Transparency and trust in what-if analysis.}
$E_C$, specializing in spatial agent-based modeling (ABM), praised the deep-learning model's ability to capture complex human behavior (in case 2) but raised the transparency issue. \textit{``ABM may oversimplify human behavior but provides experts direct access to agent attributes and states~\cite{liu2022abm}, ensuring greater transparency and control.''}  
For future projects targeting experts who prioritize rigorous experimental control, we suggest combining the strengths of deep learning with ABM, such as integrating the behavioral patterns captured by deep learning into agent-based frameworks~\cite{turgut2023framework}.

$E_D$ raised concerns about the lack of longitudinal validations for what-if analysis. Although long-term evaluation is ideal, it typically requires quasi-experimental urban data~\cite{enos2016demolition}, which is beyond the scope of our study. 
\yy{Instead, our system focuses on providing correlation-based insights for intervention suggestions.}
To this end, we integrate feature impact analysis and \yy{explicitly display CPC scores for each prediction, enabling experts to assess reliability case-by-case.}
Experts agreed that these tools enhanced their confidence by aligning the model’s reasoning with domain knowledge \yy{without requiring deep learning expertise}.
$E_A$ noted that a CPC above 60\% shows meaningful correlations that could inform their preliminary analysis, though comprehensive evaluation remains necessary.
Furthermore, $E_B$ highlighted that even these correlation-based insights, similar to those from tools like TensorFlow's What-If Tool, can suggest practical interventions when combined with expert judgment.
\yy{Future work could further enhance model trustworthiness with uncertainty visualization techniques.}

\textbf{Feasibility consideration in ``technology for social good.''}
Both $E_B$ and $E_C$ appreciated our careful consideration of the feasibility of urban planning interventions during the design phase. 
The focus on identifying areas with high potential (using TOPSIS ranking) and implementing targeted interventions was seen as a key strength for promoting a greater chance of real-world impact. 
This approach, prioritizing practical implementation, can be borrowed for other technology-for-social-good projects, helping to narrow down targets and increase feasibility.

\subsection{Limitations and Future Works}
We introduce the limitations of our system and future work.

\textbf{Lack of temporal dimension.}
Currently, \systemname only analyzes a single static snapshot of urban mobility and segregation patterns. 
However, temporal dynamics, like seasonal changes, can significantly impact segregation patterns. 
In the future, we will extend our deep learning model to analyze time-series mobility data and visualize the temporal patterns using techniques such as the Spatio-Temporal Graphs~\cite{von2015mobilitygraphs}.

\textbf{Possible bias in the demographic composition of visitors.}
We currently infer visitors' sociodemographic characteristics based on residents' census data. While this is a common practice in social science~\cite{zhang2023human, athey2021estimating}, it can introduce bias.
In the future, we aim to improve data accuracy by obtaining individual-level mobility data (\eg, telecommunication data~\cite{wu2015telcovis}) and integrating more fine-grained imputation methods, such as using housing prices~\cite{nilforoshan2023human} to infer sociodemographics.

\textbf{\yy{Scalability issues.}}
\yy{
\systemname faces both visual and computational scalability challenges as data complexity increases, which we aim to address through progressive visualization~\cite{ulmer2024progressive}. 
Visually, the \commsig and \bubbleglyph may be cluttered with numerous social groups and the \whatifview will be overwhelming with many fine-grained POI categories. 
To address this, we will explore hierarchical visualizations that let users expand or collapse categories dynamically (e.g., from broad ``food'' to more granular ``fine dining''), with meaningful aggregation levels co-designed with domain experts.
Computationally, when increased CBGs and K-nearest neighbors impact model training and what-if analysis response times, we plan to leverage process chunking, displaying intermediate results iteratively to provide immediate feedback while refining outputs progressively.
}

\section{Conclusion}
In this work, we introduce \systemname, a visual analytics system that leverages human mobility data to empower in-depth exploration and mitigation of urban segregation.
Our system combines a data analysis pipeline with an interactive interface featuring novel visualizations to support a three-level analysis workflow, enabling experts to examine segregation patterns from overview to detail and iteratively design feedback-driven urban planning interventions.
A quantitative evaluation demonstrates our model's accuracy in modeling mobility data and its capabilities for real-time interactive analysis.
Case studies and expert interviews further highlight \systemname's value in helping social scientists and policymakers uncover complex segregation patterns and design inclusive, equitable urban spaces.

\acknowledgments{
This work is partially supported by Hong Kong Research Grants Council under the Areas of Excellence Scheme grant AoE/P-601/23-N. 
}

\bibliographystyle{abbrv-doi-hyperref}
\balance
\bibliography{main}

\clearpage

\appendix
\section{Appendix}

In addition to the evaluation presented in Section~\ref{sec:model_evaluation}, we conducted a more comprehensive assessment of our adaptation strategies for the original Deep Gravity (DG) model. The evaluation source code can be found in our Github repository\footnote{\url{https://github.com/bruceyyu/inclusiviz/}} under the ``evaluation'' folder.

\subsection{Experiment Setup}

The full feature list between an origin-destination CBG pair $(v_i, v_j)$ is $concat[p_i, p_j, x_i, x_j, \pi_j, dis_{i,j}]$, where $p_i$, $p_j$ are the population sizes, $x_i$, $x_j$ represent densities of different types of POIs of the origin and destination, $\pi_j$ is a vector for the visitor characteristics of the destination, and $dis_{i,j}$ is their geographical distance.
To keep consistent with previous segregation literature, we include 12 types of POI derived from the SafeGraph dataset, including Food, Shopping, Work, Health, Religious, Service, Entertainment, Grocery, Education, Arts/Museum, Transportation, and Sports.

For each dataset, we split it into two halves: 50\% for training and 50\% for evaluation for all the models.
All models were trained for 20 epochs, repeated over five runs, to ensure stability in results.

We utilized \yy{five} widely recognized metrics to assess model performance:

\begin{itemize}
    \item \textbf{Common Part of Commuters (CPC):} We compute the Common Part of Commuters (CPC), consistent with the original work~\cite{simini2021deep}, to compare predicted flow $\hat{w}(v_i, v_j)$ with the actual $w(v_i, v_j)$, using
    \begin{equation}
        CPC(v_i, v_j) = \frac{2 \sum_{i,j} min(\hat{w}(v_i, v_j), w(v_i, v_j))}{\sum_{i,j} \hat{w}(v_i, v_j) + \sum_{i,j} w(v_i, v_j))}.
    \end{equation}
    \item \textbf{Jensen-Shannon Divergence (JSD):} JSD quantifies the difference between the predicted and actual probability distributions of commuter flows, which is defined as
    \begin{equation}
        JSD(P \parallel Q) = \frac{1}{2} D_{KL}(P \parallel M) + \frac{1}{2} D_{KL}(Q \parallel M),
    \end{equation}
    where $M = \frac{1}{2}(P + Q)$, and $D_{KL}$ is the Kullback-Leibler divergence. A lower JSD indicates a closer match between the two distributions, reflecting better model performance. 
    
    \item \textbf{Pearson Correlation:} This metric assesses the linear correlation between the predicted and actual flows. It ranges from -1 to 1, where 1 indicates a perfect positive correlation, 0 indicates no correlation, and -1 indicates a perfect negative correlation. Pearson correlation is computed as
    \begin{equation}
        r = \frac{\sum_{i,j} (w(v_i, v_j) - \bar{w})(\hat{w}(v_i, v_j) - \bar{\hat{w}})}{\sqrt{\sum_{i,j} (w(v_i, v_j) - \bar{w})^2} \sqrt{\sum_{i,j} (\hat{w}(v_i, v_j) - \bar{\hat{w}})^2}},
    \end{equation}
    where $\bar{w}$ and $\bar{\hat{w}}$ are the mean actual and predicted flows, respectively.
    
    \item \textbf{Root Mean Squared Error (RMSE):} RMSE measures the average magnitude of errors between predicted and actual flows. A lower RMSE indicates better model performance, calculated as
    \begin{equation}
        RMSE = \sqrt{\frac{1}{n} \sum_{i,j} (w(v_i, v_j) - \hat{w}(v_i, v_j))^2},
    \end{equation}
    where $n$ is the total number of flow predictions.

    \item \yy{\textbf{Normalized Root Mean Squared Error (NRMSE):} We further compute NRMSE to normalize the RMSE by the range of the actual flow $w(v_i,v_j)$, which is defined as
    \begin{equation}
        NRMSE = \frac{RMSE}{max(w(v_i,v_j) - min(w(v_i, v_j)))}.
    \end{equation}}
\end{itemize}

\subsection{Evaluation of Cities in Different Sizes}

Since Houston is a large city with around 1,500 CBGs, we also evaluated the model in a smaller city, Boston (with around 450 CBGs), using the same configurations, as shown in Table~\ref{tab:evaluation-multi-cities}.
For Houston dataset, the adapted model (DG+S+V) outperforms all other variants of the Deep Gravity model and the original Gravity model in all metrics.
For Boston dataset, the adapted model (DG+S+V) also demonstrates strong performance, surpassing the other models in all metrics.

\begin{table}[H]
    \centering
    \small
    \begin{tabular}{|l|c|c|c|c|c|}
        \hline
        Model & CPC (↑) & JSD (↓) & Pearson (↑) & RMSE (↓) & \yy{NRMSE (↓)} \\
        \hline
        \multicolumn{6}{|c|}{Houston (1506 CBGs)} \\ 
        \hline
        DG+S+V & \textbf{0.6197} & \textbf{0.3497} & \textbf{0.7733} & \textbf{200.0248} & \yy{\textbf{0.0069}} \\
        DG+V & 0.6100 & 0.3580 & 0.6775 & 218.7619 & \yy{0.0076} \\
        DG+S & 0.6125 & 0.3565 & 0.7362 & 210.4794 & \yy{0.0072} \\
        DG & 0.5979 & 0.3684 & 0.6125 & 230.5540 & \yy{0.0078} \\
        G & 0.4605 & 0.4517 & 0.3862 & 245.4852 & \yy{0.0083} \\
        \hline
        \multicolumn{6}{|c|}{Boston (462 CBGs)} \\ 
        \hline
        DG+S+V & \textbf{0.6909} & \textbf{0.2815} & \textbf{0.5874} & \textbf{61.7673} & \textbf{\yy{0.0235}} \\
        DG+V & 0.6781 & 0.2910 & 0.5629 & 65.5423 & \yy{0.0241} \\
        DG+S & 0.6829 & 0.2901 & 0.5466 & 64.3141 & \yy{0.0244} \\
        DG & 0.6809 & 0.2938 & 0.5421 & 67.1222 & \yy{0.0247} \\
        G & 0.4414 & 0.4740 & 0.3527 & 96.0198 & \yy{0.0466} \\
        \hline
    \end{tabular}
    \caption{Performance Metrics Across Models in Houston and Boston}
    \label{tab:evaluation-multi-cities}
    \vspace{-0.3cm}
\end{table}

\subsection{Evaluation of Different Population Densities}

\begin{figure}[H]
    \centering
    \begin{subfigure}[b]{0.47\linewidth}
        \centering
        \includegraphics[width=\linewidth]{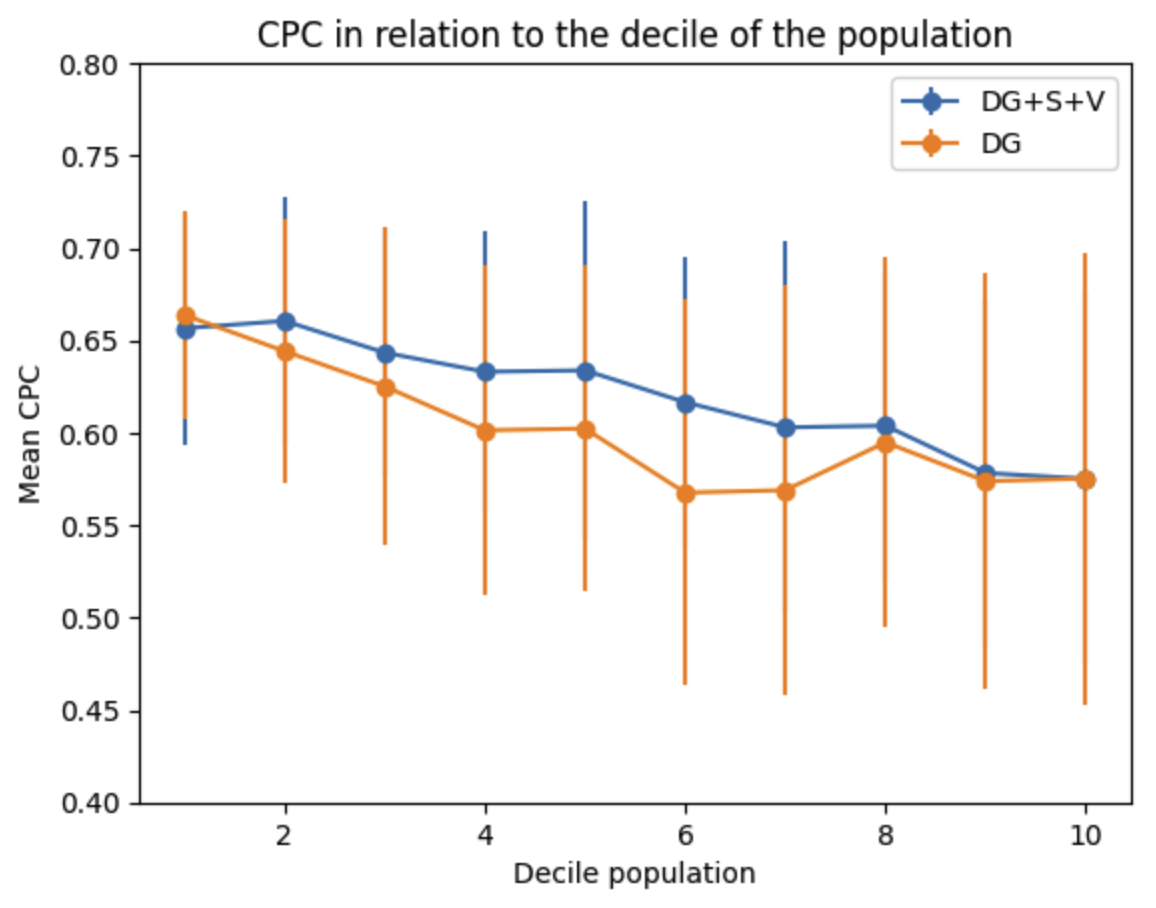}
    \end{subfigure}
    \hfill
    \begin{subfigure}[b]{0.47\linewidth}
        \centering
        \includegraphics[width=\linewidth]{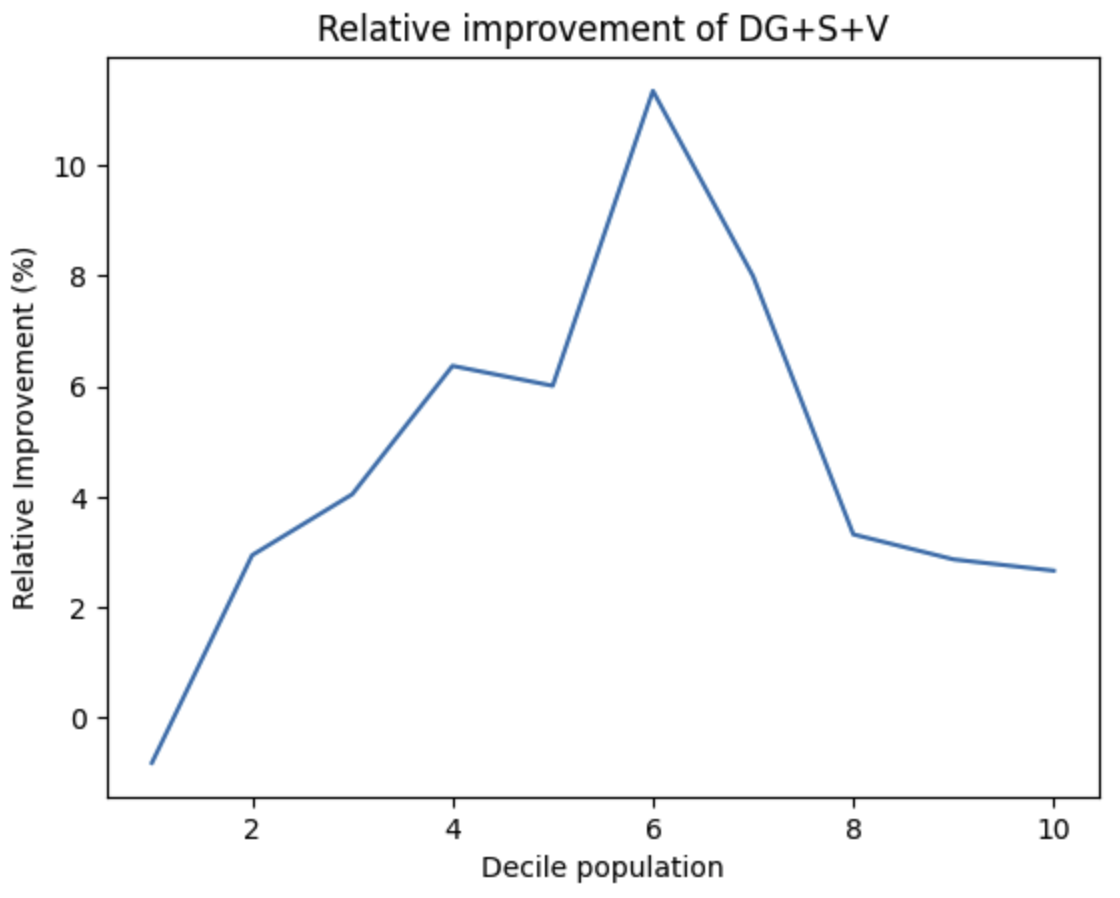}
    \end{subfigure}
    \caption{Evaluation of the Houston data: The CPC of our adapted model (DG+S+V) is comparable to the original Deep Gravity model (DG) in highly and sparsely populated regions (deciles 1 to 2, 8 to 10). In regions with moderate population sizes (deciles 3 to 7), our model shows better performance than DG.}
    \label{fig:evaluation-houston-population}
\end{figure}

We further tested the models across CBGs grouped into deciles based on population size. In Houston, the adapted model (DG+S+V) performed comparably to the original DG model in both the least (deciles 1 to 2) and most populated areas (deciles 8 to 10). However, it outperformed DG in regions with moderate population densities (deciles 3 to 7), as shown in Fig.~\ref{fig:evaluation-houston-population}.

\begin{figure}[H]
    \centering
    \begin{subfigure}[b]{0.47\linewidth}
        \centering
        \includegraphics[width=\linewidth]{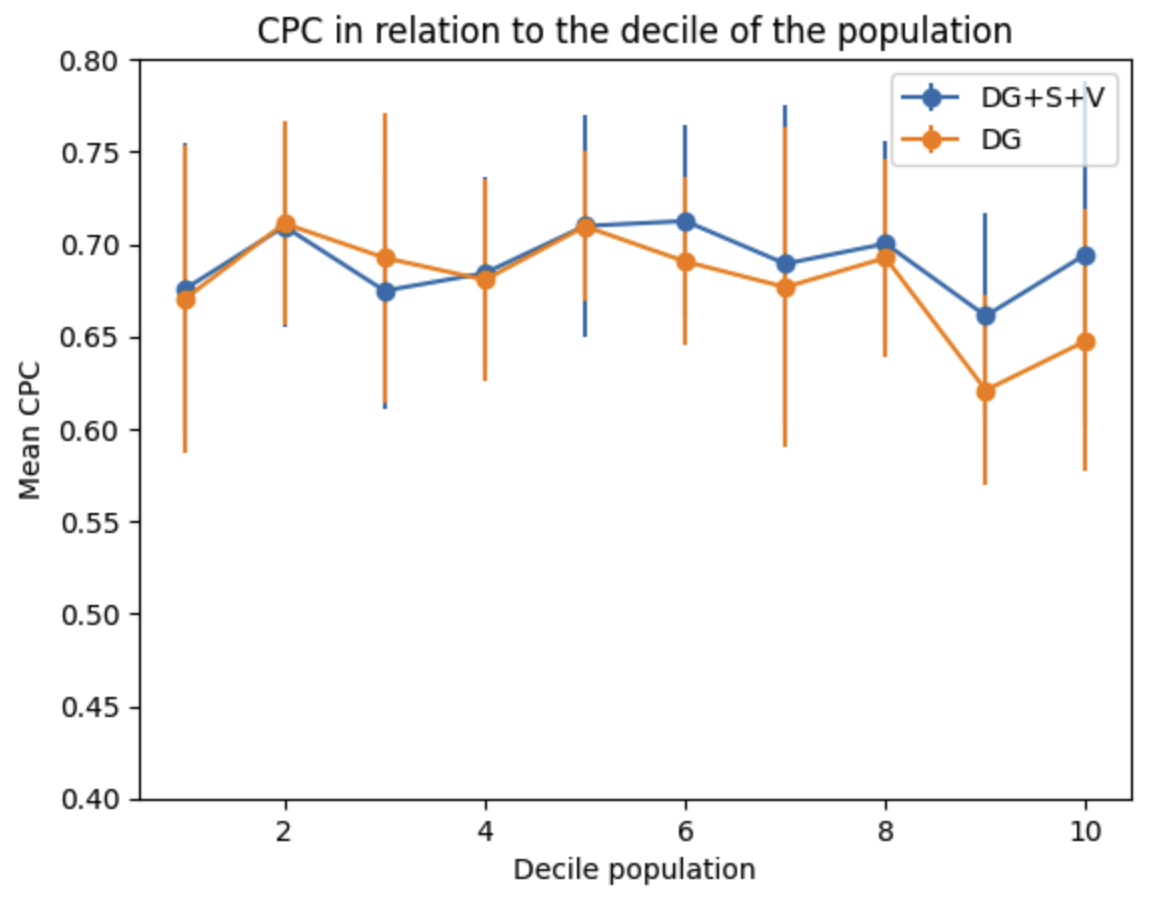}
    \end{subfigure}
    \hfill
    \begin{subfigure}[b]{0.47\linewidth}
        \centering
        \includegraphics[width=\linewidth]{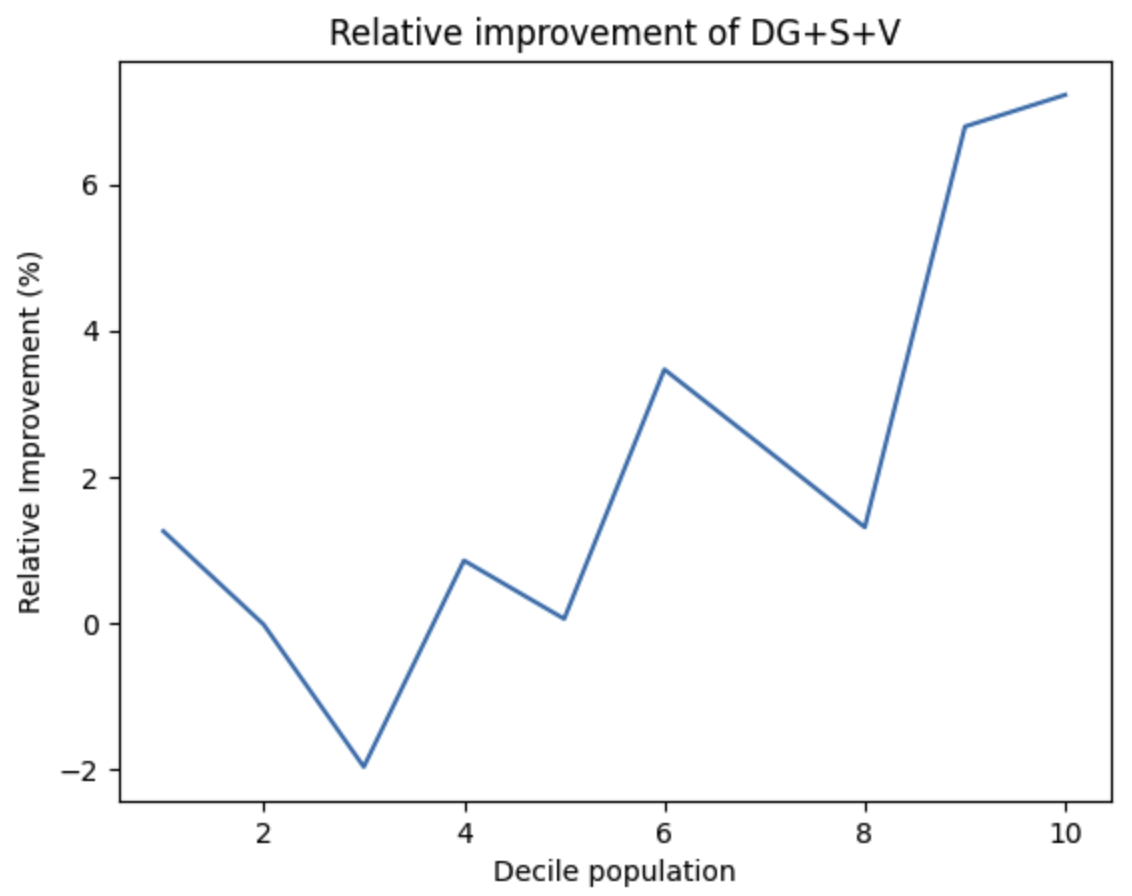}
    \end{subfigure}
    \caption{Evaluation of the Boston data: The CPC of our adapted model (DG+S+V) is comparable to the original Deep Gravity model (DG) in regions with lower population density (deciles 1 to 5). In more moderately populated regions (deciles 6 to 10), our model outperforms the original DG model.}
    \label{fig:evaluation-boston-population}
\end{figure}

In Boston, the adapted model is comparable to the DG in regions with lower population density (deciles 1 to 5), but has better performance across moderate to densely populated areas (deciles 6 to 10), as illustrated in Fig.~\ref{fig:evaluation-boston-population}.

\end{document}